\documentclass[a4paper,11pt]{article}

\usepackage{color}
\usepackage{soul}
\usepackage{amsmath,amssymb}
\usepackage{siunitx}
\usepackage{lipsum}
\usepackage{bm}
\usepackage{commath}
\usepackage{booktabs}
\usepackage{multirow}
\usepackage{rotating}
\usepackage{import}
\graphicspath{{figs/}}
\usepackage{nicefrac}
\usepackage{textcomp}
\usepackage{hyperref}

\usepackage{natbib}
\usepackage[left=25mm,right=25mm,top=1.5cm,bottom=1.5cm,includeheadfoot]{geometry}
\begin{document}

\title{Development of probabilistic dam breach model using Bayesian inference}

\author{S. J. Peter, A. Siviglia, J. Nagel,
S. Marelli,
R. M. Boes,
D. Vetsch,
and B. Sudret}
\maketitle


\abstract{
Dam breach models are commonly used to predict outflow hydrographs of potentially failing dams and are key ingredients for evaluating flood risk. In this paper a new dam breach modeling framework is introduced that shall improve the reliability of hydrograph predictions of homogeneous earthen embankment dams. Striving for a small number of parameters, the simplified physics-based model describes the processes of failing embankment dams by breach enlargement, driven by progressive surface erosion. Therein the erosion rate of dam material is modeled by empirical sediment transport formulations. Embedding the model into a Bayesian multilevel framework allows for quantitative analysis of different categories of uncertainties. To this end, data available in literature of observed peak discharge and final breach width of historical dam failures was used to perform model inversion by applying Markov Chain Monte Carlo simulation. Prior knowledge is mainly based on non-informative distribution functions. The resulting posterior distribution shows that the main source of uncertainty is a correlated subset of parameters, consisting of the residual error term and the epistemic term quantifying the breach erosion rate. The prediction intervals of peak discharge and final breach width are congruent with values known from literature. To finally predict the outflow hydrograph for real case applications, an alternative residual model was formulated that assumes perfect data and a perfect model. The fully probabilistic fashion of hydrograph prediction has the potential to improve the adequate risk management of downstream flooding.
}


\section{Introduction}

Earthen dams have been built by humans to store water for multiple purposes for millennia \citep{Schnitter_1994}. While they are regarded as safe structures, history taught us that they nevertheless may fail. Extensive literature is available reporting historic dam failures \citep[e.g.][]{Broich_1996}. To protect people and infrastructure downstream of a potentially failing dam, today's supervising authorities and dam operators take precautionary measures, install emergency warning systems, and set up evacuation plans. To do so, reliable predictions about the amount and timing of released water in case of a dam failure are needed. This information comes from dam break models which provide outflow hydrographs for subsequent flood routing models.

In this paper the focus is on homogeneous and non-cohesive earthfill embankment dams. Processes that belong to the breach formation are of interest here, whereas any analysis concerning probability of failure, breach initiation, or flood propagation downstream is not considered. Herein the main breach formation process is regarded as the enlargement of an initial breach due to slope erosion of dam material leading to increasing breach discharge. Research has put much effort into understanding the complex natural phenomena of breach formation processes, both by field tests \citep[e.g.][]{Morris_2005} and laboratory experiments \citep[e.g.][]{Schmocker_2012, Frank_2016}. The gained insights help understand the breach formation process and provide relevant information for model development. However, their application to real-life breach situations is often limited due to simplifications and scale effects \citep{ASCE_2011}. Regardless of the complexity and high non-linearity of dam breach processes, the necessity of dam breach prediction tools induced the development of a multitude of numerical models over the last decades. The physical properties usually investigated are the outflow hydrograph and the size of the final breach. Dam breach modeling techniques vary strongly in complexity and can be classified according to their level of detail:
\begin{enumerate}
\item \textit{Statistically-based} models are purely data-driven and rely on regression analysis of historical dam breach events and make estimations about embankment breach characteristics, e.g. peak discharge, time to failure, or final breach width. The dam breach model thereby can be described either by power functions of governing dam-reservoir quantities \citep{Froehlich_2008} or by applying methods of artificial intelligence \citep{Amini_2011, Hooshyaripor_2012}. Since no physics are considered in these models, their reliability and explanatory power is generally low.
\item \textit{Simplified physics-based} models take into account the description of selected physical processes, e.g. draw-down of the reservoir and enlargement of the breach over time due to erosion of dam material. Because of their simplicity, many a-priori assumptions have to be made, e.g. initial breach size, breach geometry, flow over the breach, sediment transport, and geomechanical concepts. The computational cost of these models is low and the number of parameters that have to be defined is commonly small, hence allowing for real time applications \citep{Ma_2012}. A comprehensive list of simplified dam breach models can be found in \citet{ASCE_2011}.
\item \textit{Detailed physics-based} models describe the embankment breaching process in one, two, or even three dimensions by applying sophisticated numerical approaches. During the last decades these models became popular due to advances in computer sciences \citep[e.g.][]{Broich_1996,Wang_2006, Faeh_2007, Wu_2009, Volz_2013}. Nevertheless, describing and parameterizing not only the macroscopic but also the microscopic phenomena of a gradually failing dam is very challenging. In particular, the interaction between water and dam material inside the breach, the short term stability of saturated and compacted soils, and the quantification of high-concentration sediment transport capacity pose difficulties.
\end{enumerate}

Before dam breach models are applied as prediction tools, no matter what type of model, they are often calibrated to a data set. Historical dam breach data are therefore used to feed the models, but data on real-life embankment failures are usually poorly documented for various reasons \citep{ASCE_2011}. Uncertainties on breach properties, such as average breach width and failure time, have been reported in \citet{Froehlich_2008}, and their relation to outflow hydrograph are investigated in \cite{Ahmadisharaf_2016}. Uncertainties of dam breach models are often quantified in terms of prediction errors, which are minimized during model calibration by selecting appropriate data and fitting procedures. To the best of our knowledge, full quantification of uncertainties in physical dam and reservoir properties and output variables has not been explored in dam breach modeling. In simplified physics-based models systematic sensitivity analysis has been carried out \citep{Fread_1984,Walder_1997,Delorenzo_2014}, unlike in detailed models where the computational cost is too high. Surrogate modeling techniques, also known as meta-modeling or response surface modeling, could offer an alternative to this end, such as Kriging \citep{Santner_2003} and polynomial chaos expansion \citep{Xiu_2002}. \citet{Wahl_2004} assessed the prediction errors of various statistically-based embankment breach models by applying them to a set of 108 dam failures. The resulting prediction intervals are approximately $\pm 1/3$, and $\pm 1/2$ to $1$ orders of magnitude for predicted breach width and peak outflow, respectively. In 2013 the International Committee of Large Dams (ICOLD) organized a numerical benchmark where the participants were invited to predict the breach outflow hydrograph of a hypothetical dam failure, i.e. the true outcome was not known \citep{Graz_2013}. The results demonstrate that the variations between the predicted hydrographs of different dam breach models/modelers are large and can significantly influence the result of hydrodynamic calculations \citep{Escuder_2016}. The origin of the discrepancies in the resulting hydrographs of the benchmark participants can be mainly attributed to variability of the erodibility of embankment material as a result of different soil types, compaction effort, and water content \citep{Morris_2008}. Ultimately, on the basis of a single model the wide range of possible breach outflow hydrographs cannot be quantified reliably enough for prediction purposes.

On these grounds many researchers emphasized the need for a systematic quantification of all types of uncertainties in dam breach models to reliably predict embankment dam failure processes \citep{Wahl_2004, Froehlich_2008, ASCE_2011} and consequently to incorporate into comprehensive risk management systems \citep{Altinakar_2009}. This has been achieved e.g. in hydrological sciences by first promoting the use of uncertainty estimation as routine \citep{Pappenberger_2006}, setting up an integrated risk management framework \citep{Buechele_2006}, critically discussing the deterministic and probabilistic approaches \citep{Baldassarre_2010}, and proposing new methodologies within the probabilistic framework \citep{Alfonso_2016}.

Generally, uncertainties are categorized as (i) aleatory, that describe the natural variability of a physical process, (ii) epistemic, that describe the lack of knowledge in parametric description of the process, and (iii) global model inadequacy and data uncertainty \citep{Kennedy_2001}. Based on this classification \citet{Nagel_2016} proposed a unified framework to quantify uncertainties on all levels by using Bayesian inverse modeling and a deterministic model as backbone. Related frameworks were applied in different fields, such as 3-D environment modeling \citep{Balakrishnan_2003}, sediment entrainment modeling \citep{WuChen_2009}, groundwater modeling \citep{Laloy_2012,Shi_2014}, rating curve derivation \citep{Mansanarez_2016}, or design flood estimation \citep{Steinbakk_2016}.

The goal of the present paper is to enhance the reliability of hydrograph predictions by developing a new dam break model in a fully probabilistic manner. The model is herein regarded as a strong approximation of an open system. Accordingly the verification and validation of the truth of such a model is nearly impossible \citep{Oreskes_1994}. However, a model can be conditionally confirmed, similar to the original idea of Rev. Thomas Bayes: for given evidence in certain circumstances, find the model, out of a set of feasible models, that explains the evidence best. The model predictions will supposedly be more reliable (i) the more evidence is consulted to conditionally confirm the model, (ii) the higher the evidence quality is, (iii) the more widely accepted physical knowledge is implemented in the model, (iv) the more sophisticated methods are applied to find the best model.

In this vein, a new deterministic dam breach model is proposed that is embedded into a Bayesian multilevel framework \citep{Nagel_2016}. Therein the vector of model parameters is split into different parameter types according to their uncertainty character. The underlying parameter distributions are defined by Bayesian inference, in which the prior information is specified by empirical knowledge and the reference data is based on measured quantities of dam failures. The underlying population of the data is represented by worldwide and historically failed, man-made, homogeneous embankment dams. The novelty comprises a complete quantification of parametric as well as residual uncertainties in dam breach modeling. The latter consists of both data and structural model errors. The outcome of the present study is not only the final probabilistic dam breach model, but instead providing a framework that allows the integration of further data. Since the underlying deterministic model is treated as a black-box, it can be replaced in a simple manner for future analysis. As stated by \citet{Morris_2008}, it is not easy to improve the accuracy of dam breach models, that is the degree to which model predictions represent a real dam failure. Nevertheless, by incorporating the information about model uncertainties these predictions become more reliable, i.e. they are more trustworthy in case of not knowing the true outcome of a possible dam failure. The enhanced reliability of predicted breach hydrographs is of major importance regarding the quality of risk quantification induced by dam failures.

This paper is organized as follows: The deterministic dam breach model formulation and its parameters is introduced in Section~\ref{sec:model}. The outline of the Bayesian multilevel approach is given in Section~\ref{sec:calibration}, including successfully embedding the dam breach model into the probabilistic framework. The resulting quantification of uncertainties of all levels is shown in Section~\ref{sec:results}. Finally critical points are discussed within an examplary model application in Section~\ref{sec:discussion}and conclusions are drawn in Section~\ref{sec:conclusions}.


\section{Dam Breach Model}\label{sec:model}

The aim of this section is to present the development of the deterministic dam breach model, which is the backbone of the probabilistic framework proposed in this paper. The model under investigation can be attributed to simplified physics-based dam breach models. Because of prevailing lack of data in dam breach modeling, the number of parameters is kept as small as possible to circumvent the risk of too many degrees of freedom. In addition it is of paramount importance to have a computationally efficient model due to its probabilistic inversion where million evaluations are needed. At the same time the model must be able to reproduce the main dam breach phenomenon. The physical basis of the breach formation process is the water-sediment interaction: the discharging water is the driving force of breach erosion, and vice versa the breach enlargement controls the rate of discharge \citep[e.g.][]{Singh_1988_1}.

\subsection{Formulation of the Physical Processes}

Processes that lead to an initial breach are not considered here. In reality, the initial breach is being formed by a multitude of different failure causes, e.g. overtopping, internal erosion, slope instabilities, or foundation problems \citep{Foster_2000}. This breach initiation process is much slower than the subsequent breach enlargement and varies strongly between different failure causes. Regardless, discharge rates are still low and do not influence the hydrograph of a failing dam \citep{Wahl_2004,Morris_2008}. After developing a sufficiently large breach, the flow rate increases rapidly and the failure cannot be prevented.

\begin{figure}[ht]
\centerline{\includegraphics[height=3in]{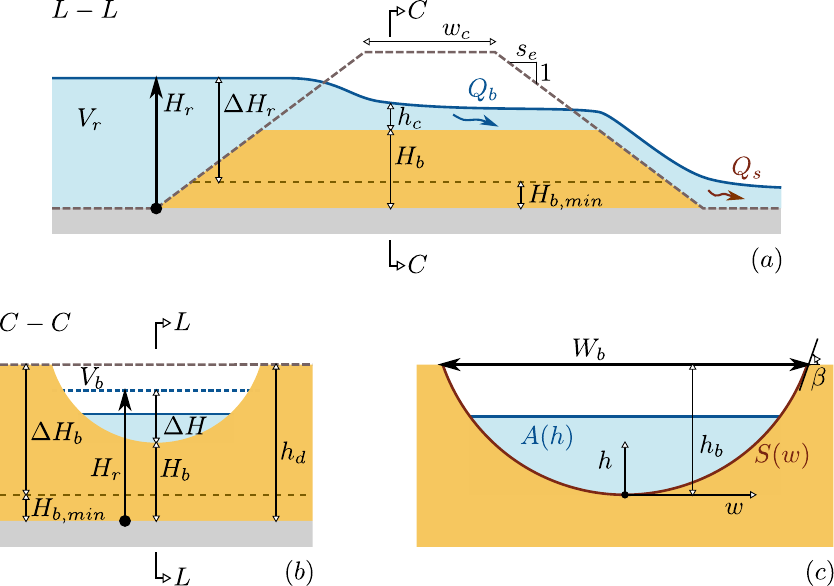}}
\caption{Longitudinal (a) section $L-L$ and transverse (b) control section $C-C$ of the dam-reservoir system, and (c) breach shape definition. Thick arrows indicate the unknown variables $H_r$ and $W_b$ in Eq.~\eqref{eq:formation}.}
\label{fig:sketches}
\end{figure}

This breach formation process is essentially dominated by the mechanics of overtopping \citep{Singh_1996, ASCE_2011, Delorenzo_2014}) and according to literature the main physical processes involved are (i) gradual erosion of dam material, (ii) breach enlargement, (iii) increasing breach outflow, and (iv) decreasing water level in the reservoir. These four processes can generally be formulated by a system of two ordinary differential equations (ODE) (see notation in Figure\ref{fig:sketches})
\begin{subequations}
\label{eq:formation}
\begin{align}
\label{eq:formation_Hr}
\dod{H_r}{t} &= -Q_b \del{\dod{V_r}{H_r}}^{-1}\\[1em]
\label{eq:formation_Wb}
\dod{W_b}{t} &= Q_s \del{\dod{V_b}{W_b}}^{-1}.
\end{align}
\end{subequations}
Eq.~\eqref{eq:formation_Hr} represents a continuity equation for conserving the water volume stored in the reservoir $V_r \sbr{\si{\cubic\metre}}$, where the depletion of the reservoir level $H_r \sbr{\si{\metre}}$ over time $t \sbr{\si{\second}}$ is described by the discharge of breach outflow $Q_b \sbr{\si{\cubic\metre\per\second}}$ and the rate of change of reservoir volume with respect to reservoir level $\od{V_r}{H_r} \sbr{\si{\square\metre}}$. The continuity equation Eq.~\eqref{eq:formation_Wb} is conserving the volume of eroded dam material $V_b \sbr{\si{\cubic\metre}}$, where the increasing breach width $W_b \sbr{\si{\metre}}$ over time is controlled by the discharge of dam material that is transported out of the breach $Q_s \sbr{\si{\cubic\metre\per\second}}$ and the rate of change of breach volume with respect to breach width $\od{V_b}{W_b} \sbr{\si{\square\metre}}$. Similar formulations have been adopted by \citet{Singh_1988_1} and \citet{Macchione_2008a}. In the model presented here, processes that belong to the mechanism of progressive surface erosion are considered only, whereas the mechanisms of head cutting (cohesive dam material) and interlocking (coarse dam material) are neglected. Dams showing cohesive or rockfill materials are clearly outside of the application range of this model.

\begin{figure}[ht]
\centerline{\includegraphics[height=1in]{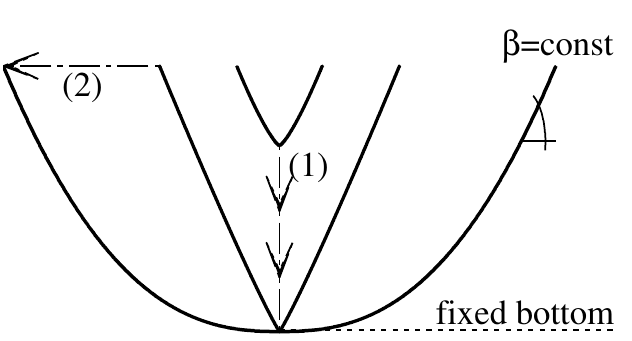}}
\caption{Breach shape development with constant breach side angle $\beta$ at the top and fixed bottom ($H_{b,min}$): stage of vertical erosion (1), i.e. from initial breach to fixed bottom ({$H_b>H_{b,min}$}); and stage of lateral widening (2), i.e. no further deepening of the breach ($H_b=H_{b,min}$). In case of low erodibility the stage of vertical erosion will never be exceeded, what can be referred to as partial failure.}
\label{fig:shape}
\end{figure}

As depicted in Figure~\ref{fig:shape}, the breach enlargement is split into two distinct breach development states in time. In a first stage the breach development is dominated by vertical erosion. The breach width $W_b$ is enlarged and breach bottom level $H_b \sbr{\si{\metre}}$ is lowered at the same time with constant rate, preserving self-similarity of the breach shape over time \citep{Pickert_2011,Frank_2016}. When the breach bottom reaches the foundation of the dam $H_{b,min} \sbr{\si{\metre}}$ it is assumed that no further deepening of the breach is possible and lateral widening is controlling the breach enlargement only \citep[e.g.][]{Coleman_2002, Chinnarasri_2004}. Thus the development of the breach bottom is described by the following simple ODE:
\begin{equation}
\od{H_b}{W_b} = \left\{
\begin{array}{ll}
-\frac{h_b}{W_b} & \textrm{if} \quad H_b>H_{b,min},\\[1em]
0 & \textrm{if} \quad H_b=H_{b,min},
\end{array}\right.
\label{eq:dHbdWb}
\end{equation}
where $h_b = h_d-H_b \sbr{\si{\metre}}$ is the breach height and $h_d \sbr{\si{\metre}}$ is the dam height (see Figure~\ref{fig:sketches}).

Despite reducing the complex physical processes of breach formation into simple continuity equations \eqref{eq:formation_Hr} and \eqref{eq:formation_Wb}, it has been shown that the physical processes contained in this are sufficient to reproduce the hydrograph of rather detailed models \citep{Vonwiller_2015} or laboratory experiments \citep{Chinnarasri_2009}. Below further details are given about assumptions that are made in this study to parameterize and quantify the variables of the right-hand side in the system of equations~(\ref{eq:formation}).

\subsection{Breach Geometry}

In this section the simplifications of the breach geometry description are stated. The final goal is to quantify the breach volume change rate $\od{V_b}{W_b}$ (see Eq.~\eqref{eq:formation_Wb}), which is dependent on geometrical parameters only.

The longitudinal breach shape is reminiscent of an hourglass shape and the hydraulic control section is defined by the curved weir crest at the inlet of the breach \citep{Walder_2015}. In order to lower the complexity of the model and to reduce the number of model parameters, the three dimensional breach geometry is simplified as a prismatic channel. Its cross-section is regarded as the hydraulic control section (see Eq.~\eqref{eq:control_section}). For different existing parameter models the description of the cross-sectional shape is varying, i.e. triangular, trapezoidal, rectangular, or parabolic \citep{ASCE_2011}. To include all types of shapes here, the breach area is described as a power law of $h$, i.e. $A \propto h^k$ (see Figure~\ref{fig:sketches}c), where $k$ is a shape parameter ranging from $k=1$ representing a rectangular breach shape to $k=2$ representing a triangular shape. Strictly speaking, the model requires $k<1$ and the breach shape converges to a rectangular shape for $k\to1$. A similar approach describing breach shapes of instantaneous dam breaks is followed by \citet{Pilotti_2010}. For given top width $W_b$ and bottom level $H_b$ the breach side wall is obtained as
\begin{equation}
S\del{w} = h_b \del{\frac{2|w|}{W_b}} ^ \frac{1}{k-1},
\label{eq:Sx}
\end{equation}
where $w \in \intcc{-\frac{W_b}{2},\frac{W_b}{2}}$ is a control variable running across the breach section, starting at the breach center (see Figure~\ref{fig:sketches}c). The length of the breach wall on one side in the interval $\sbr{a,b}$ is
\begin{equation}
S_L\del{a,b} = \int\limits_a^b\!\sqrt{1 + S^{\prime^2}}\,\dif{w},
\label{eq:Sl}
\end{equation}
where
\begin{equation}
S^\prime\del{w} = \frac{h_b}{k-1}\del{\frac{2}{W_b}w^{2-k}}^\frac{1}{k-1}
\label{eq:Sprime}
\end{equation}
is the breach side slope (for $w\geq0$). This is not calculated analytically and has to be integrated numerically. Further, for given water level $h$ inside the breach, the water surface width $W$ and the corresponding breach area $A$ are
\begin{equation}
W\del{h} = W_b\del{\frac{h}{h_b}}^{k-1},
\quad
A\del{h} = \frac{W_b}{k h_b^{k-1}} h^k.
\label{eq:WA}
\end{equation}
The breach volume is calculated by integration of $A\del{h}$ along the breach (i.e. across the dam) with $w_c$ being the crest width and $s_e$ the embankment slope of the dam, hence
\begin{equation}
V_b = \frac{W_b h_b}{k} \del{w_c + \frac{2 s_e h_b}{k+1}}.
\end{equation}

The breach side angle $\beta$ at the top of the breach (see Figure~\ref{fig:sketches}c and Figure~\ref{fig:shape}) represents the short-term critical failure angle of the dam material and is assumed to be constant during the erosion process \citep{Chinnarasri_2004,Frank_2016}. This angle can exceed the long-term critical failure angle by far even for non-cohesive soil materials, due to stabilizing effects of the apparent cohesion \citep{Volz_2013,Volz_2017}. Setting the breach side slope at the top $S^\prime\del{W_b/2}=\tan\beta$ (see Eq.~\eqref{eq:Sprime}), the shape exponent is then given by
\begin{equation}
k = \frac{2h_b}{W_b\tan\beta}+1.
\label{eq:k}
\end{equation}

The breach volume rate of change $\od{V_b}{W_b}$ in Eq.~\eqref{eq:formation_Wb} is different for the two breach development states vertical erosion $H_b>H_{b,min}$ and lateral widening $H_b=H_{b,min}$. In the first case the breach depth grows steadily with increasing top width (see Figure~\ref{fig:sketches} and Eq.~\eqref{eq:dHbdWb}) and therefore both $\od{h_b}{W_b}$ and $k$ are constant. After the breach bottom has reached the dam foundation, the breach depth does not change anymore, but $k$ is now decreasing with enlarged breach (see Eq.~\eqref{eq:k}). After some algebraic calculations the breach volume rate of change is finally given by
\begin{equation}
\od{V_b}{W_b} =
\left\{\begin{array}{lr}
h_b \sbr{ \frac{2}{k} w_c + \frac{6}{k^2\del{k+1}} s_e h_b }
& \textrm{if} \quad H_b>H_{b,min},\\[1em]
h_b \sbr{ \frac{2k-1}{k^2} w_c + \frac{2\del{3k^2-1}}{k^2\del{k+1}^2} s_e h_b }
& \textrm{if} \quad H_b=H_{b,min}.
\end{array}\right.
\label{eq:Vb_Wb}
\end{equation}

\subsection{Breach Hydraulics}\label{sec:hydraulics}

In this section all hydraulic variables inside the breach are defined, including information about the underlying assumptions. Finally the breach outflow $Q_b$ to be used in Eq.~\eqref{eq:formation_Hr} is estimated.

The hydraulic variables are defined within the prismatic channel where the occurrence of critical flow (Froude number $Fr = 1$) is assumed to hold (see Figure~\ref{fig:sketches}a). The effect of the streamline curvature, implicating an energy head loss and $Fr < 1$ \citep{Walder_2015}, is not included here. This effect could be incorporated by introducing an additional parameter, e.g. discharge coefficient, or the Froude number. The breach geometry of this control section is assumed to be representative for the entire breach. At the same time it acts only as a hydraulic control section and therefore the location of this transition in flow regime is not relevant for the model \citep{Singh_1996,Coleman_2002,Macchione_2008a}. The energy head is given by
\begin{equation}
H_e = H_r - H_b = h + \frac{v^2}{2g},
\label{eq:control_section}
\end{equation}
where $h$ is the water depth, $v$ is the average flow velocity within the breach and $g$ is the acceleration due to gravity. The critical flow condition implies the maximum breach discharge for given $H_e$. Thus the critical water depth and flow velocity are
\begin{equation}
h_c = \frac{2k}{2k+1} H_e, \qquad v_c = \sqrt{g\frac{h_c}{k}}.
\label{eq:hc_vc}
\end{equation}
Consequently the breach discharge can be written as
\begin{equation}
Q_b = A\del{h_c} v_c = \frac{W_b}{h_b^{k-1}} \sqrt{\frac{g}{k^3}} {h_c}^{k+\nicefrac{1}{2}}.
\label{eq:Qb}
\end{equation}
In addition the hydraulic radius of the critical cross-section ($h=h_c$) can be calculated through the following relation
\begin{equation}
r_h = \frac{A\del{h}}{P_w\del{h}},
\quad
P_w\del{h} = 2 \cdot S_L\del{0,\frac{W\del{h}}{2}}
\label{eq:rhy}
\end{equation}
where $P_w\del{h}$ is the wetted perimeter along the breach side walls up to level $h$ (see Eq.~\eqref{eq:Sl} and Eq.~\eqref{eq:WA}).

\subsection{Breach Erosion}\label{sec:erosion}

In this section the parameters that control the sediment transport rate inside the breach are introduced. Ultimately the sediment transport rate out of the breach $Q_s$ is defined, which is needed in Eq.~\eqref{eq:formation_Wb}.

The flow velocity $v$ and the corresponding hydraulic radius $r_h$ are supposed to control the sediment transport within the breach. Empirical sediment transport formulas that quantify bed, suspension or total load are commonly formulated as $q_s \propto \del{\tau_b-\tau_c}^{c_1} v^{c_2} r_h^{c_3} \sbr{\si{\square\metre\per\second}}$, where $\tau_b \propto r_h J$ is the bottom shear stress for steady flow conditions and $\tau_c$ is the critical shear stress for incipient motion. The breach formation process is regarded as intense bedload phenomena, where $\tau_b >> \tau_c$, and the threshold $\tau_c$ is neglected therefore. The energy slope $J$ can be defined by applying empirical flow formulas of type $v \propto r_h^{c_4} J^{0.5}$. Exponents $c_{1\ldots4}$ are constants that differ between distinct empirical formulas. Knowing the flow velocity $v$ from Eq.~\eqref{eq:hc_vc} and hydraulic radius $r_h$ from Eq.~\eqref{eq:rhy})in the control section of the breach, the transport rate is then
\begin{equation}
q_s = \gamma \cdot v^{\nu} \cdot {r_h}^{\eta}
\label{eq:qs}
\end{equation}
with $\gamma$ being a global scaling coefficient that acts as a tuning factor to accelerate or hinder the erosion process, and $\nu = 2c_1+c_2$ and $\eta = c_1\del{1-2c_4}+c_3$ are exponents that combine the empirical formulation of transport and friction laws. For example \citet{Macchione_2008a} applied the sediment transport equation by Meyer-Peter and M\"{u}ller ($c_1=\frac{3}{2}, c_2=c_3=0$) and the friction law by Strickler ($c_4=\frac{2}{3}$), which yields $\nu=3$ and $\eta=-0.5$. The two exponents are physically interpreted as: (i) the larger $\nu \del{> 0}$ is, the stronger will be the influence of the hydraulic condition on the sediment transport (high flow velocity leads to more erosion); and (ii) the smaller $\eta \del{< 0}$ is, the stronger will be the influence of breach geometry on the sediment transport (narrow shapes leads to more erosion).

The rate of sediment that is transported out of the breach is finally quantified as
\begin{equation}
Q_s = P_e \cdot q_s,
\quad
P_e = 2\cdot S_L\del{w_0,\frac{W\del{h}}{2}},
\label{eq:Qs}
\end{equation}
where $P_e$ is the erodible perimeter, defined by the length along the breach side wall $S_L$ bounded by
\begin{equation}
w_0 = \left\{
\begin{array}{ll}
0 & \textrm{if} \quad H_b>H_{b,min},\\[1em]
\frac{2-k}{k}\frac{W\del{h}}{2} & \textrm{if} \quad H_b=H_{b,min},
\end{array}\right.
\end{equation}
and the transverse location where the water surface intersects with the breach side. Hence in case of vertical erosion the erodible perimeter is equal to the wetted perimeter. During lateral widening the erodible perimeter is dependent on the breach shape: for triangular shape ($k=2$) the complete wetted perimeter is accessible for sediment transport, for rectangular shape ($k\to1$) only the vertical part of the breach is erodible, and for all intermediate breach shapes a smooth transition between these two extremes occurs, i.e. the closer the breach shape is to a rectangle, the smaller will the erodible zone get due to fixed dam foundation in the middle of the breach.

\subsection{Reservoir Depletion}

The aim of this section is to quantify the reservoir volume change rate $\od{V_r}{H_r}$. It is the last missing variable for the characterization of the ODE in Eq.~\eqref{eq:formation}.

The reservoir retention curve is parameterized by a power function
\begin{equation}
V_r = V_{r,0} \del{\frac{H_r}{H_{r,0}}}^\alpha,
\end{equation}
where $V_{r,0}$ and $H_{r,0}$ are initial values for the water volume and its associated reservoir level, e.g. the storage volume and full supply level. The exponent $\alpha$ characterizes the reservoir topography, i.e. $\alpha=1.0$ represents a rectangular basin whereas $\alpha=4.0$ denotes a basin located in rather mountainous valleys \citep{Kuehne_1978}. Consequently, the reservoir volume depletion rate can be described as
\begin{equation}
\od{V_r}{H_r} = \frac{\alpha V_{r,0}}{{H_{r,0}}^\alpha} {H_r}^{\alpha-1}.
\label{eq:Vr_Hr}
\end{equation}

\subsection{Summary and Numerical Solution of the Parametric Breach Model}

To solve the system of ODEs in Eq.~\eqref{eq:formation}, the geometrical properties of the dam-reservoir system ($h_d$, $w_c$, $s_e$, $\beta$, $\alpha$) have to be defined. Furthermore the state variables of the system and their initial values have to be set given all model parameters (see Table~\ref{tbl:parameters}): the drop in reservoir level $\Delta H_r$, released reservoir volume $\Delta V_r$, and final breach height $\Delta H_b$. Detailed information are provided in Appendix~\ref{app:IC}.

In each time step, the depletion rate of the reservoir level $\od{H_r}{t}$ and the widening rate $\od{W_b}{t}$ are calculated through Eqs.~\eqref{eq:Vb_Wb}, \eqref{eq:Qb}, \eqref{eq:Qs}, and \eqref{eq:Vr_Hr}. The system of ODEs is numerically integrated using classic four step Runge-Kutta scheme, initially choosing a large time step. By re-integration with decreased time step, the peak of the resulting breach outflow hydrograph will be closer to its exact value. This iteration is done until a certain relative precision is reached. In this study a value of $1 \textrm{\textperthousand}$ was chosen.

\begin{table}[t]
\caption{Parameter vector $\bm{x}$ of the deterministic dam breach model, categorized by their physical meaning.}
\centering
\begin{tabular}{l l}
\toprule
description & $\bm{x}$\\
\midrule
\emph{dam properties:} & \\
dam height [\si{\metre}] & $h_d$ \\
crest width [\si{\metre}] & $w_c$ \\
embankment slope [$-$] & $s_e$ \\
\midrule
\emph{reservoir properties:} & \\
reservoir basin shape [$-$] & $\alpha$ \\
reservoir level drop [\si{\metre}] & $\Delta H_r$ \\
released reservoir volume [\si{\cubic\metre}] & $\Delta V_r$ \\
\midrule
\emph{breach properties:} & \\
breach side angle [\si{\degree}] & $\beta$ \\
final breach height [\si{\metre}] & $\Delta H_b$ \\
initial breach depth relative to $\Delta H_r$ [-] & $r_0$ \\
\midrule
\emph{erosion properties:} & \\
exponent for flow velocity [$-$] & $\nu$ \\
exponent for hydraulic radius [$-$] & $\eta$ \\
scaling coefficient for transport rate [$-$] & $\gamma$ \\
\bottomrule
\end{tabular}
\label{tbl:parameters}
\end{table}


\section{Probabilistic Model Calibration}\label{sec:calibration}

The deterministic dam breach model outlined in the previous section is hereinafter treated as 'black box' where no use of information about the mathematical model implemented by the numerical code is made \citep{Kennedy_2001}. Generally, the calibration procedure described in this section may be applied without fundamental modifications to other dam breach models than the one previously introduced, varying in level of details and/or parameterization. However, overly detailed models with many parameters significantly increase the risk of over-parametrization, especially when only scarce or poor-quality data is available. The model of concern is formally described as a function
\begin{equation}
\bm{\tilde{y}} = \mathcal{M}\del{\bm{x}},
\label{eq:model}
\end{equation}
where the vector of model parameters $\bm{x}$ is mapped to the vector of model outputs $\bm{\tilde{y}}$. The parameters $\bm{x}$ are listed in Table~\ref{tbl:parameters}. The model outputs of interest in this study are
\begin{equation}
\bm{\tilde{y}} = \del{\log_{10}\tilde{Q}_p,\log_{10}\tilde{W}_f},
\end{equation}
with $\tilde{Q}_p \sbr{\si{\cubic\meter\per\second}}$ being the peak discharge of the outflow hydrograph and $\tilde{W}_f = \nicefrac{\tilde{W}_b}{k} \sbr{\si{\meter}}$ the average breach width at the end of the breaching process. According to literature, where errors of these quantities are given as orders of magnitude \citep{Wahl_2004,ASCE_2011}, a logarithmic scale is chosen here.

When applying the proposed dam breach model to predict a hydrograph of an existing but potentially failing embankment dam, some model parameters in $\bm{x}$ can be easily quantified (e.g. crest width $w_c$), while others are not known due to lack of data (e.g. basin reservoir shape $\alpha$) or they are unobservable and act as tuning coefficients (e.g. scaling coefficient of sediment transport $\gamma$). Parameters that show a strong epistemic character, e.g. parameters that quantify the breach erosion (see Eq.~\eqref{eq:qs}), will undergo a Bayesian update in the present study. These parameters are defined as vector $\bm{\phi}$ and called \textit{Quantities of Interest} (\textit{QoI}) hereafter. The goal is to draw conclusions about the epistemic model inputs from real observation data $\langle \bm{y}_i \rangle$. Henceforward $\langle \bm{y}_i \rangle$ stands for a sequence $\langle \bm{y}_i \rangle_{1 \leq i \leq n} = \del{\bm{y}_1,\bm{y}_2,\ldots,\bm{y}_n}$ of $n$ experiments, containing $Q_p$ and $W_f$ of historically failed embankment dams (see Table~\ref{tbl:data}). Applying Bayes' rule yields the joint posterior density
\begin{equation}
\pi\del{\bm{\phi} \mid \langle \bm{y}_i \rangle} =
\frac{\mathcal{L}\del{\bm{\phi};\langle \bm{y}_i \rangle} \cdot \pi\del{\bm{\phi}}}{C},
\label{eq:posterior}
\end{equation}
where $\mathcal{L}\del{\bm{\phi};\langle \bm{y}_i \rangle}$ is the likelihood function, $\pi\del{\bm{\phi}}$ the prior distribution of the model parameters, and $C$ is a normalization constant such that the integral of the posterior distribution $\pi\del{\bm{\phi} \mid \langle \bm{y}_i \rangle}$ is equal to one.

The unknown parameters in $\bm{\phi}$ are subsequently quantified by Bayesian inference, that is ``the process of fitting a probability model to a set of data and summarizing the result by a probability distribution on the parameters of the model and on observed quantities such as predictions for new observations'' \citep{Gelman_2014}. This is achieved by performing the following steps of classical Bayesian data analysis:
\begin{enumerate}
\item \textit{Setting up a full probabilistic model}, i.e. the previously defined dam breach model is put into a Bayesian multilevel framework to assess different levels of uncertainty according to the underlying physical problem. This includes the specification of prior knowledge and the formulation of the likelihood function.
\item \textit{Conditioning the model on observed data}, i.e. setting up a residual model that describes the forward model discrepancy, quantifying the prior knowledge, defining the likelihood function with given data of historical dam failures, and finally calculating the appropriate posterior distribution of the unobserved quantities of interest.
\item \textit{Evaluating the goodness-of-fit}, i.e. analyzing the posterior distribution and compare the output of the now calibrated model with the observed data.
\end{enumerate}

These three steps are further detailed in the next sections. Further steps would consist of including additional data not considered yet for model conditioning, i.e. performing the three steps again whereas the posterior distribution now represents the new prior knowledge.

\subsection{Bayesian Multilevel Framework}\label{sec:BMLM}

\begin{figure}[ht]
\centerline{\includegraphics[height=2.2in]{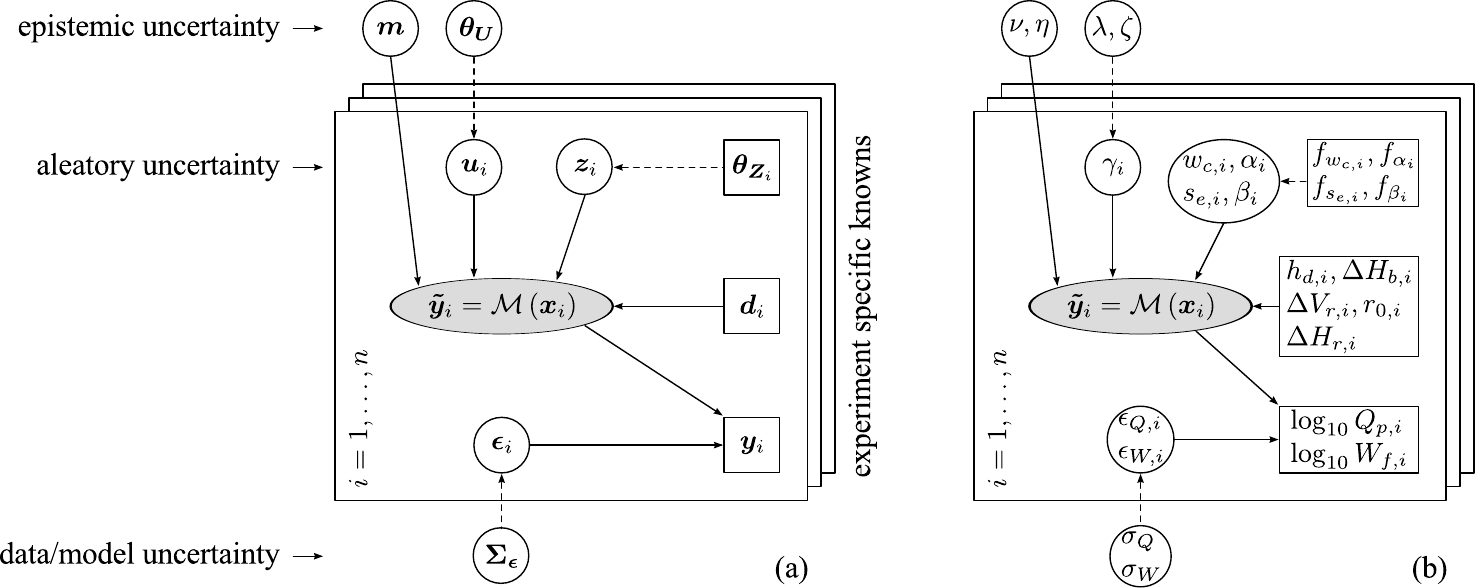}}
\caption{Directed acyclic graph (DAG) representing the probabilistic multilevel modeling approach followed in this study: vertices represent known ($\bm{\square}$) or unknown ($\bm{\circ}$) quantities, whereas directed edges symbolize their deterministic ($\rightarrow$) or probabilistic ($\dashrightarrow$) relations (adapted from \cite{Nagel_2016}). On the left side the general formulation is shown (a), whereas on the right side the actual parameters of the dam breach model and their role in the multilevel framework are shown (b).}
\label{fig:bmlm}
\end{figure}

The probabilistic shell that is shaped around the deterministic dam breach model can be represented by a Bayesian network (see Figure~\ref{fig:bmlm}). The Bayesian multilevel approach applied herein is adopted from \cite{Nagel_2016}. It provides ``a natural framework for solving complex inverse problems in the presence of natural variability and epistemic uncertainty''. The multilevel character of the method at hand is given by the hierarchically composed sub-models, such as the deterministic forward model itself, different categories of parameter uncertainty and/or variability described by the prior model, and prediction errors of the forward model specified in the residual model.

\begin{figure}[ht]
\centerline{\includegraphics[height=1.5in]{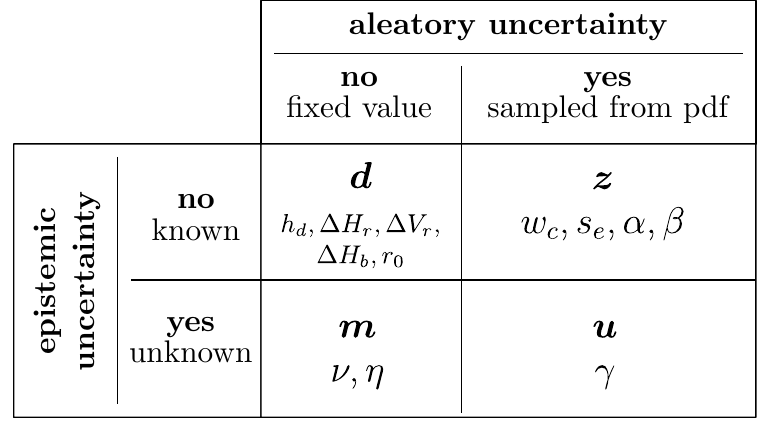}}
\caption{Classification of the dam breach model parameters according to their (un)certain nature: the combinations of aleatory and epistemic uncertainty result in four different classes. The parameters with known values and known probability distribution function (pdf) ($\bm{d}$, $\bm{z}$ respectively) are experiment specific whereas the the parameters with unknown values and unknown hyper-parameters ($\bm{m}$, $\bm{u}$ respectively) are assumed to be fixed for all experiments (see Figure~\ref{fig:bmlm}).}
\label{fig:classification}
\end{figure}

The parameters in $\bm{x}$ of the forward model (see Eq.~\eqref{eq:model}) can be categorized according to their physical meaning (see Table~\ref{tbl:parameters}). Alternatively they can be classified according to four different categories of parameters (see Figure~\ref{fig:classification})
\begin{equation}
\bm{x} = \del{\bm{m},\bm{u},\bm{z},\bm{d}}
\label{eq:x_multilevel}
\end{equation}
that differ in their (un)certain nature, as done in \cite{Eicher_2014}. This classification is based on general data availability on the one hand, and quality of empirical parameter knowledge on the other hand. The classification proposed in this study is one possible solution. Under different circumstances and/or available expert knowledge the result might be an alternative classification.

The first category is defined as well-known experimental conditions $\bm{d} = \del{h_d, \Delta H_r, \Delta V_r, \Delta H_b, r_0}$, that are normally reported for historical dam failure events. These quantities are assumed to be perfectly known for each experiment $i$ and they can be considered as deterministic arguments of the dam breach model (see Table~\ref{tbl:data}). The second category of parameters $\bm{z} = \del{w_c, s_e, \alpha, \beta}$ are subject to known and experiment-specific aleatory uncertainty. Their true values are not known as for $\bm{d}$ but follow the probability distributions $f_{\bm{Z}}\del{\bm{z}_i;\bm{\theta_Z}}$. That is, the hyper-parameters $\bm{\theta_Z}$ are assumed to be well-known, whereas the vector of realizations $\bm{z}_i$ throughout the experiments $i=1 \ldots n$ is not known. Parameters belonging to this class are usually not reported for historical dam failures but there is data available from which we can define the population distribution function \citep{Eicher_2014}. Furthermore, parameters that are subject to epistemic uncertainty $\bm{m} = \del{\nu, \eta}$, i.e. exponents of the transport rate formula in Eq.~\eqref{eq:qs}, are treated as constant model parameters, but their true values are not known. Ultimately, $\bm{u}$ represents unknown parameters that are subject to aleatory uncertainty that itself is unknown and not changing across the experiments. In case of the dam breach model at hand the scaling coefficient $\gamma$ controlling the velocity of dam material erosion via the sediment transport in Eq.~\eqref{eq:qs} is the only parameter that belongs to this category. Thereby not only the realizations are unknown, but as well the hyperparameters $\bm{\theta_U}$ describing the probability function $f_{\bm{U}}\del{\bm{u}_i;\bm{\theta_U}}$ from which realizations $\bm{u}_i$ are drawn. Therein contained is the parametric uncertainty of the erosion process itself. However, the distribution family of $f_{\bm{U}}$ is assumed to be known.

\begin{sidewaystable}
\caption{Data and its experiment-specific knowns applied in this study for model inversion, originating from \citet{Wahl_1998}: The underlying population of the data are worldwide, historically failed, homogeneous earthfill embankment dams. Existing and potentially failing dams belonging to this population may be investigated with the model proposed in this study. Experimental conditions, uncertain parameters and observations are listed. Where good data about $s_{e}$ and $w_{c}$ are available, the location parameter (mean) is defined based on this information and the scale parameter (standard deviation) is chosen to be much smaller than the location. Otherwise $s_e$ is drawn from $N_{s_e} = N\del{2.16,0.66}$ bounded by $\sbr{1,10}$, and $w_c$ from $LN_{w_c} = LN\del{1.55,0.51}$, respectively \citep{Eicher_2014}. Due to missing information the realization of $\alpha$ and $\beta$ are sampled from $U_{\alpha} = U\del{1,4}$ and $U_{\beta} = U\del{45,90}$, respectively, representing the physically meaningful range.}
\centering
\footnotesize{
\begin{tabular}{r l
*{1}{S[table-format=2.2]} *{3}{S[table-format=2.3]} *{1}{S[table-format=1.1]}
*{4}c
*{1}{S[table-format=4.0]} *{1}{S[table-format=3.0]}
}

\toprule

\multicolumn{2}{c}{Failure} &
\multicolumn{5}{c}{experimental conditions $\bm{d}_i$} &
\multicolumn{4}{c}{uncertain parameters $\bm{z}_i$} &
\multicolumn{2}{c}{observations $\bm{y}_i$}\\

\cmidrule(lr){1-2}
\cmidrule(lr){3-7}
\cmidrule(lr){8-11}
\cmidrule(lr){12-13}

 \(i\) &
 Name &
 \(h_d\) [$\si{\metre}$] &
 \(\Delta V_r\) [$\SI{e3}{\cubic\metre}$] &
 \(\Delta H_r\) [$\si{\metre}$] &
 \(\Delta H_b\) [$\si{\metre}$] &
 \(r_0\) [$-$] &
 \(f_{s_e}\) [$\si{\metre}$] &
 \(f_{w_c}\) [$\si{\metre}$] &
 \(f_\alpha\) [$-$] &
 \(f_\beta\) [$\si{\degree}$] &
 \(Q_p\) [$\si{\cubic\metre\per\second}$] &
 \(W_f\) [$\si{\metre}$]\\

\midrule

1 & Apisapha &
34.1 & 22.2 & 28.0 & 31.1 & 0.2 &
\(N\del{2.5,0.01}\) & \(LN\del{1.59,0.01}\) & \(U_{\alpha}\) & \(U_{\beta}\) &
6850 & 93\\

2 & Baldwin Hills &
71.0 & 0.910 & 12.2 & 21.3 & 0.6 &
\(N\del{1.9,0.01}\) & \(LN\del{2.95,0.01}\) & \(U_{\alpha}\) & \(U_{\beta}\) &
1130 & 25\\

3 & Butler &
7.16$^{b}$ & 2.38 & 7.16 & 7.16$^{a}$ & 0.2 &
\(N_{s_e}\) & \(LN_{w_c}\) & \(U_{\alpha}\) & \(U_{\beta}\) &
810 & 63\\

4 & Fred Burr &
10.4 & 0.750 & 10.2 & 10.4 & 0.2 &
\(N_{s_e}\) & \(LN_{w_c}\) & \(U_{\alpha}\) & \(U_{\beta}\) &
654 & \hrulefill \\

5 & French Landing &
12.2 & 3.87 & 8.53 & 12.2 & 0.2 &
\(N\del{2.25,0.01}\) & \(LN\del{0.88,0.01}\) & \(U_{\alpha}\) & \(U_{\beta}\) &
929 & 27\\

6 & Frenchman Creek &
12.5 & 16.0 & 10.8 & 12.5 & 0.2 &
\(N\del{2.5,0.01}\) &  \(LN\del{1.81,0.01}\) & \(U_{\alpha}\) & \(U_{\beta}\) &
1420 & 55\\

7 & Hatchtown &
19.2 & 14.8 & 16.8 & 18.3 & 0.2 &
\(N\del{2.25,0.01}\) & \(LN\del{1.81,0.01}\) & \(U_{\alpha}\) & \(U_{\beta}\) &
3080 & 151\\

8 & Ireland no. 5 &
5.18$^{b}$ & 0.160 & 3.81 & 5.18 & 0.2 &
\(N_{s_e}\) & \(LN\del{0.88,0.01}\) & \(U_{\alpha}\) & \(U_{\beta}\) &
110 & 14\\

9 & Johnstown &
38.1 & 18.9 & 24.6 & 24.6 & 0.2 &
\(N\del{1.75,0.01}\) & \(LN\del{1.12,0.01}\) & \(U_{\alpha}\) & \(U_{\beta}\) &
8500 & 95\\

10 & Lawn Lake &
7.9 & 0.798 & 6.71 & 7.62 & 0.2 &
\(N\del{1.55,0.01}\) &  \(LN\del{0.88,0.01}\) & \(U_{\alpha}\) & \(U_{\beta}\) &
510$^{c}$ & 22\\

11 & Lily Lake &
3.66$^{b}$ & 0.093 & 3.35 & 3.66 & 0.2 &
\(N_{s_e}\) & \(LN_{w_c}\) & \(U_{\alpha}\) & \(U_{\beta}\) &
71 & 11\\

12 & Little Deer Creek &
26.2 & 1.36 & 22.9 & 26.2 & 0.6 &
\(N_{s_e}\) & \(LN\del{1.81,0.01}\) & \(U_{\alpha}\) & \(U_{\beta}\) &
1330 & 30\\

13 & Lower Latham &
7.01 & 7.08 & 5.79 & 7.01 & 0.2 &
\(N_{s_e}\) & \(LN\del{1.53,0.01}\) & \(U_{\alpha}\) & \(U_{\beta}\) &
340 & 79\\

14 & Propsect &
4.42$^{b}$ & 3.54 & 1.68 & 4.42 & 0.2 &
\(N_{s_e}\) & \(LN\del{1.46,0.01}\) & \(U_{\alpha}\) & \(U_{\beta}\) &
116 & 88\\

15 & Quail Creek &
21.3$^{b}$ & 30.8 & 16.7 & 21.3 & 0.2 &
\(N_{s_e}\) & \(LN_{w_c}\) & \(U_{\alpha}\) & \(U_{\beta}\) &
3110 & 70\\

\bottomrule
\multicolumn{13}{l}{$^{a}$ Assumed value: final breach height $\Delta H_b$ is equal to drop in reservoir level $\Delta H_r$.}\\[-0.5em]
\multicolumn{13}{l}{$^{b}$ Assumed value: dam height $h_d$ is equal to final breach height $\Delta H_b$.}\\[-0.5em]
\multicolumn{13}{l}{$^{c}$ Value obtained by numerical simulation.}\\

\end{tabular}
}
\label{tbl:data}
\end{sidewaystable}

\subsection{Residual Model}\label{sec:residual}

Even when assuming perfectly known model parameters $\del{\bm{m},\bm{u}_i,\bm{z}_i,\bm{d}_i}$, predictions $\bm{\tilde{y}}_i$ are expected to deviate from real observations $\bm{y}_i$ due to the output imperfection $\bm{\epsilon}$, that can be regarded as a container of measurement errors, numerical approximations, and general model inadequacies. In that sense observations
\begin{equation}
\bm{y} = \bm{\tilde{y}} + \bm{\epsilon}
\end{equation}
are interpreted as outcomes of a random process and the underlying random generating mechanism, the joint probability distribution of the model parameters, is investigated. Imperfections $\bm{\epsilon}$ are hereinafter referred to as residuals and are assumed to be realizations of a Gaussian random variable $\bm{E} \sim f_{\bm{E}} \del{\bm{\epsilon};\bm{\Sigma_\epsilon}}$, usually centered around $\bm{0}$ with variance $\bm{\Sigma_\epsilon}$. Errors due to numerical approximations and model inadequacies are collected as structural errors herein. To separate them from imperfections $\bm{\epsilon}$, more sophisticated residual error models could be used, e.g. representing the structural errors by a functional error term \citep{Brynjarsdottir_2014}. This analysis is not included in this study because of missing information about dam breach data uncertainties and because it is out of the scope of this study.

The unknown error of observed quantities $Q_p$ and $W_f$ and of model inadequacies is taken into account during the calibration procedure, referred to as \textit{residual calibration}. For simplicity uncorrelated and normally distributed errors are assumed with variances
\begin{equation}
\bm{\Sigma_\epsilon} = \sbr{
\begin{array}{cc}{\sigma_Q}^2 & 0 \\ 0 & {\sigma_W}^2\end{array}
}.
\label{eq:sigma_epsilon}
\end{equation}
The standard deviations of both the error in peak discharge ${\sigma_Q} \sbr{\si{\log_{10}\cubic\metre\per\second}}$ and final breach width ${\sigma_W} \sbr{\si{\log_{10}\metre}}$ are collected in the vector $\bm{\sigma} = \del{\sigma_Q,\sigma_W}$ such that $\bm{\Sigma_\epsilon} = \bm{\sigma}^2 \bm{I}_2$ and will be quantified through inference analysis.

\subsection{Prior Knowledge}\label{sec:priors}

In a Bayesian fashion prior or expert knowledge is seen as a subjective degree of belief about the true values of the parameters. Prior parameter knowledge is formulated by means of probability distribution functions. Through Bayesian data analysis this prior knowledge is updated with a set of data resulting in posterior parameter distributions. It is distinguished between (1) structural priors and (2) parametric priors.

The former include information about the prior model of experiment-specific unknowns $\bm{z}_i$ and $\bm{u}_i$. It is referred to as prescribed uncertainty because the corresponding variabilities are integrated out in the marginalized formulation of the likelihood function (more details in Section~\ref{sec:likelihood}). Therefore, this type of prior knowledge cannot be improved by additional experiments. The structural priors in this study are formulated as distribution functions $f_{\bm{Z}}\del{\bm{z}_i;\bm{\theta}_{\bm{Z}_i}}$ to draw samples $\bm{z}_i$ of the parameter vector $\bm{z}$ (see Table~\ref{tbl:data}) and
\begin{equation}
f_{\bm{U}}\del{\bm{u}_i;\bm{\theta_U}} = LN\del{\gamma_i;\lambda,\zeta}
\label{eq:fU}
\end{equation}
being a lognormal distribution to draw samples of the scaling coefficient $\gamma_i$. The hyperparameters $\bm{\theta_U} = \del{\lambda,\zeta}$ are unknowns associated with parametric priors (see below) and composed by the location parameter $\lambda$ (mean value of the associated normal distribution) and the scale parameter $\zeta > 0$ (standard deviation of the associated normal distribution).

Second, the parametric priors describe the knowledge about the global unknowns $\bm{m}$, $\bm{\theta_U}$, and $\bm{\sigma}$. Model parameters that belong to one of the aforementioned categories are herein referred to as quantities of interest \textit{QoI} during Bayesian inferential analysis and collected as the vector $\bm{\phi}=\del{\bm{m},\bm{\theta_U},\bm{\sigma}}$. Their prior formulation will be updated when Bayesian data analysis is applied. In case of $\bm{m} = \del{\nu,\eta}$ the expert knowledge falls back on empirical formulas quantifying sediment transport and hydraulic friction (see \appref{app:transport}) and are herein defined as bivariate normal distribution
\begin{equation}
\pi_{\bm{M}}\del{\bm{m}} \sim \bm{N}_2\del{\bm{\mu_m},\bm{\Sigma_m}}
\label{eq:prior_m}
\end{equation}
with $\bm{\mu_m}$ consisting of mean values $\mu_{\nu}=4.0$ and $\mu_{\eta}=-0.5$, and $\bm{\Sigma_m}$ being the covariance matrix with $\sigma_{\nu}=0.9$, $\sigma_{\eta}=0.3$, and $\rho_{\nu\eta}=-0.1$. The hyperparameters $\bm{\theta_U}=\del{\lambda,\zeta}$ that quantify the structural prior of the scaling coefficient $\gamma_i$ cannot be guessed based neither on expert nor on empirical knowledge. This is due to lack of data or even unobservability. Therefore the prior
\begin{equation}
\pi_{\bm{\Theta_U}}\del{\bm{\theta_U}} =
\begin{pmatrix}\pi_{\Lambda} \\ \pi_{Z}\end{pmatrix} \sim
\begin{pmatrix}U\del{-15,5} \\ U\del{0,2}\end{pmatrix}
\label{eq:prior_gamma}
\end{equation}
does not contain any valuable information apart from physically feasible and broad enough ranges. The prior knowledge about structural model and data uncertainty constituted in random variable $\bm{\epsilon}$ is represented by non-informative uniform distributions
\begin{equation}
\pi_{\bm{E}}\del{\bm{\Sigma_\epsilon}} =
\begin{pmatrix}\pi_{\sigma_Q} \\ \pi_{\sigma_W}\end{pmatrix} \sim
\begin{pmatrix}U\del{0.0,0.6} \\ U\del{0.0,0.6}\end{pmatrix}.
\label{eq:prior_epsilon}
\end{equation}
Finally, the joint prior distribution is defined as a function of the quantities of interest
\begin{equation}
\pi\del{\bm{\phi}} = \pi\del{\bm{m},\bm{\theta_U},\bm{\Sigma_\epsilon}} = \pi_{\bm{M}}\del{\bm{m}} \cdot \pi_{\bm{\Theta_U}}\del{\bm{\theta_U}} \cdot \pi_{\bm{E}}\del{\bm{\Sigma_\epsilon}}
\label{eq:prior}
\end{equation}
and can be applied in Bayes' formula to evaluate the posterior distribution (see Eq.~\eqref{eq:posterior}).

\subsection{Definition of the Likelihood Function}\label{sec:likelihood}

The probability of observed data $\bm{y}_i$ for given model parameters $\del{\bm{m},\bm{\theta_U}}$ and residual variability $\bm{\Sigma_\epsilon}$ is defined as the likelihood function
\begin{equation}
\mathcal{L}\del{\bm{m},\bm{\theta_U},\bm{\Sigma_\epsilon};\langle \bm{y}_i \rangle}
= \prod\limits_{i=1}^{n} f\del{\bm{y}_i \mid \bm{m},\bm{\theta_U},\bm{\Sigma_\epsilon}}.
\label{eq:likelihood}
\end{equation}
The data $\bm{y}_i$ used in this study are listed in Table~\ref{tbl:data}. The likelihood depends on the experiment specific knowns $\del{\bm{\theta}_{\bm{Z}_i},\bm{d}_i}$ and is conditional on the unknowns $\del{\bm{m},\bm{\theta_U},\bm{\Sigma_\epsilon}}$. Here a marginalized (or integrated) formulation
\begin{equation}
\begin{aligned}
f\del{\bm{y}_i \mid \bm{m},\bm{\theta_U},\bm{\Sigma_\epsilon}} = &\int_{\mathcal{D}_u} \int_{\mathcal{D}_z} f_{\bm{E}}\del{\bm{y}_i - \mathcal{M}\del{\bm{m},\bm{u}_i,\bm{z}_i,\bm{d}_i}; \bm{\Sigma_\epsilon}}\\
&f_{\bm{U}\mid\bm{\Theta_U}}\del{\bm{u}_i\mid\bm{\theta_U}} f_{\bm{Z}\mid\bm{\Theta_Z}}\del{\bm{z}_i\mid\bm{\theta_Z}} \dif{\bm{u}_i} \dif{\bm{z}_i}
\end{aligned}
\label{eq:likelihood_int}
\end{equation}
is used, wherein the aleatory uncertainty in the so-called latent variables $\del{\bm{u},\bm{z}}$ is integrated out \citep{Nagel_2016}. This formulation focuses on global unknown parameters and therefore can be seen as a function of quantities of interest $\bm{\phi}=\del{\bm{m},\bm{\theta_U},\bm{\Sigma_\epsilon}}$. Alternatively the latent variables $\del{\bm{u},\bm{z}}$ can be inferred, avoiding the numerical integration of Eq.~\ref{eq:likelihood_int}. However, the knowledge about future realizations of $\del{\bm{u},\bm{z}}$ cannot be improved thereby.

The formal likelihood function $f\del{\bm{y}_i \mid \bm{m},\bm{\theta_U},\bm{\Sigma_\epsilon}}$ of data set $i$ is numerically approximated by Monte-Carlo integration. Through independently sampling of $\del{\bm{u}_i^{\del{k}},\bm{z}_i^{\del{k}}}$ from their population distributions and further calculating the vector of residuals $\bm{\epsilon}_i^{\del{k}} = \bm{y}_i - \mathcal{M}\del{\bm{m},\bm{u}_i^{\del{k}},\bm{z}_i^{\del{k}},\bm{d}_i}$ the approximation of the formal likelihood function is determined. The distribution function $f_{\bm{E}}$ of the residuals, described by the covariance matrix $\bm{\Sigma_\epsilon}$ (see Eq.~\eqref{eq:sigma_epsilon}), is applied directly and the likelihood is therefore estimated as
\begin{equation}
\hat{f}\del{\bm{y}_i\mid\bm{m},\bm{\theta_u},\bm{\Sigma_\epsilon}} = \frac{1}{K}\sum\limits_{k=1}^{K}f_{\bm{E}}\del{\bm{\epsilon}_i^{\del{k}};\bm{\Sigma_\epsilon}}.
\end{equation}
During the process of likelihood estimation the number of model evaluations $K$ is incrementally increased until a relative precision on the estimated likelihood $\hat{f}\del{\bm{y}_i \mid \bm{m},\bm{\theta_U},\bm{\Sigma_\epsilon}}$ of $1\si{\percent}$ is reached. Usually $K=10^4$ model evaluations are needed to reach this target.

\subsection{Model Inversion}\label{sec:MCMC}

In a deterministic framework inverse modeling connotes finding the parameter set which produces the model output that fits best the observed data. In a Bayesian world inverse modeling implies updating the prior knowledge with observed data to end up with the posterior distribution. In addition to this full uncertainty picture, point estimates of the posterior, like mean or mode, are chosen as characterization of the high dimensional distribution in practice. With $\bm{\phi} = \del{\bm{m},\bm{\theta_U},\bm{\Sigma_\epsilon}}$ being the vector containing the \textit{QoI} of the multilevel framework introduced in Eq.~\eqref{eq:x_multilevel}, the joint posterior distribution in Eq.~\eqref{eq:posterior} is recalled as
\begin{equation}
\pi\del{\bm{\phi} \mid \langle \bm{y}_i \rangle} \propto
\mathcal{L}\del{\bm{\phi} ; \langle \bm{y}_i \rangle} \cdot
\pi\del{\bm{\phi}}.
\label{eq:posterior_mcmc}
\end{equation}
To perform inferential analysis, random samples are needed from the posterior distribution. Even if the $d$-dimensional posterior is well defined in the space of the unknown parameters $d=\mathcal{D}_{\bm{\phi}}=\mathcal{D}_{\bm{m}} \times \mathcal{D}_{\bm{\theta_U}} \times \mathcal{D}_{\bm{\Sigma_\epsilon}}$, drawing samples from it is not trivial nonetheless. The standard approach to circumvent this challenge is the method of MCMC. In this study a differential evolution Markov chain (DE-MC) algorithm is applied \citep{Braak_2006} (see Appendix~\ref{app:mcmc_alg} for more details). Thereby generated samples of the target distribution, i.e. the joint posterior distribution in Eq.~\eqref{eq:posterior_mcmc}, can be further processed to finalize inference analysis and interpreting the impact on the breach model.


\section{Inference Analysis Results}\label{sec:results}

Statistical inference of the proposed dam breach model will be presented here. First, the methodology of how to interpret the results of inference analysis is introduced. Second, the results of the model inversion are demonstrated with $\bm{\phi}=\del{\lambda,\zeta,\nu,\eta,\sigma_Q,\sigma_W}$. Third, the results are discussed, especially with the focus on real case applications, where results of additional analysis is presented.

\subsection{Methodology of Interpretation}

MCMC simulation was run with the choice of population size $N = 12$, satisfying $N=2*d$ for unimodal target distribution \citep{Braak_2006}. Details concerning the MCMC performance are given in Appendix~\ref{app:mcmc_perf}. The total number of model evaluations during MCMC simulation was approximately $I \cdot N \cdot n \cdot K \approx 5\cdot10^9$. An efficient implementation of deterministic dam breach model and highly parallel execution yields $10^4$ model evaluations per second, resulting in a total run time of roughly $150 \si{\hour}$. 

Since the initial motivation of this study was to improve the reliability of dam breach model predictions, the goodness-of-fit (GoF) of the presented parameter model is discussed. For this purpose the model is run using the same input data as used for model inversion, including the structural priors and the updated parametric prior, i.e. the posterior of \textit{QoI}. The most probable parameter set $\hat{\bm{\phi}}$ of the $d$-dimensional joint posterior is taken to run the dam breach model. Hence, the variability $\tilde{\bm{y}}$ originates purely from aleatory uncertainty. Goodness-of-fit evaluations are done by visually comparing the predicted quantities $\tilde{\bm{y}}$ with the observed data $\bm{y}$ in form of violin plots \citep{Hintze_1998}. Further the GoF statistics
\begin{equation}
\bm{r} = \tilde{\bm{y}} - \bm{y} + \bm{\epsilon}
\label{eq:gof_residual}
\end{equation}
are processed in a descriptive manner. The expected value $\widehat{\textrm{E}}\sbr{\bm{r}}$ is calculated for each data $i$ and afterwards averaged over the whole sequence of $n$ data sets to estimate potential prediction biases. Furthermore, the sequence $\langle\bm{p}_i\rangle$ with $\bm{p}_i=\text{Pr}\sbr{\bm{r}_i\leq0}$ being the percentile of data $i$ is presented in form of a percentile plot. This way the location of potential prediction biases is visualized. As it is done for existing dam breach models \citep{Wahl_2004}, the $95\%$ interval $\widehat{\textrm{I}}_{95}\sbr{\bm{r}} \approx 2.0\sqrt{\widehat{\textrm{Var}}\sbr{\bm{r}}}$ is estimated to quantify the width of prediction band, where $\widehat{\textrm{Var}}\sbr{\bm{r}}$ is the sample variance of GoF statistics $\bm{r}$. All analysis is done for overall objectives $\bm{y}$ and separately for both objectives peak discharge $Q_p$ and final breach width $W_f$. This is because the model inversion is performed including the information of both objectives and any potential bias in predicting solely e.g. $Q_p$ must be prevented.

Comparing the GoF statistics simplifies the interpretation of the resulting model accuracy and reliability and the implication on the application to predict a potential dam breach hydrograph. Quantitative summaries of this analysis are found in Table~\ref{tbl:mode} and Table~\ref{tbl:error}.

\subsection{Posterior Distribution}

\begin{figure}[t]
\centerline{\includegraphics[height=3in]{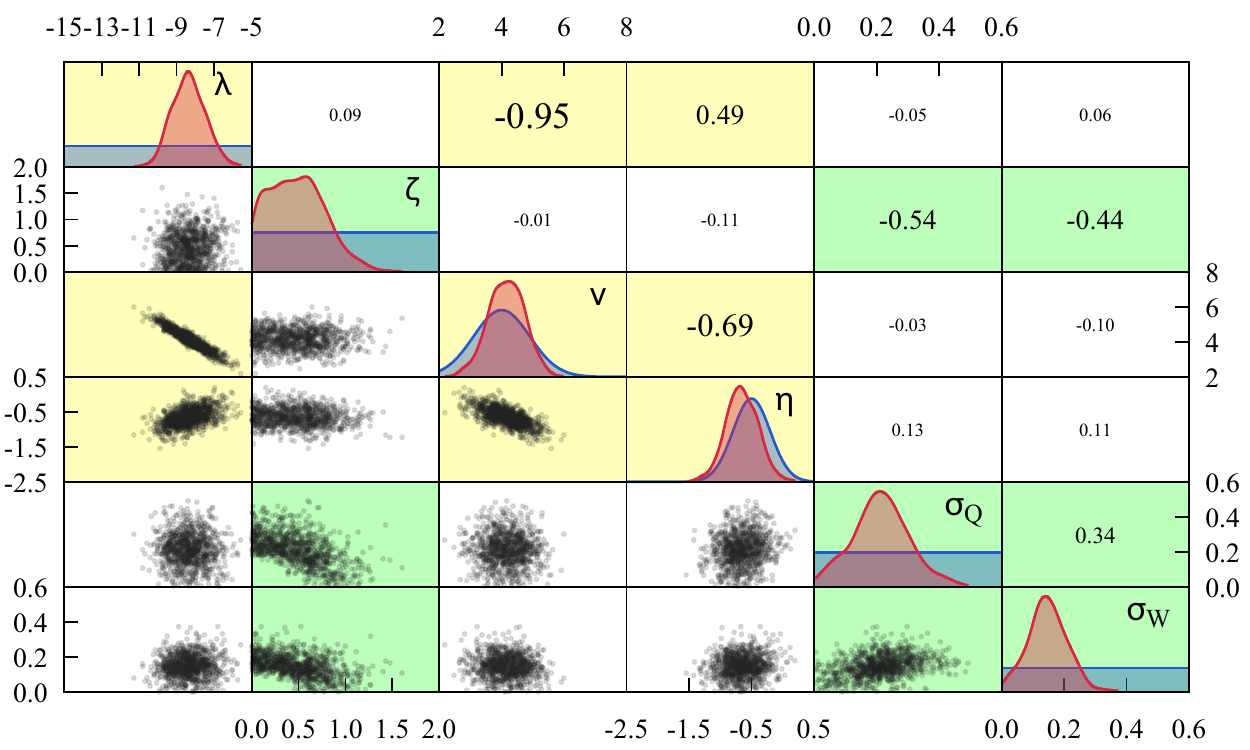}}
\caption{Posterior distribution samples: the diagonals show the marginals, posterior (red) as well as prior (blue) for purposes of comparison; the upper right part indicates the correlation coefficient between parameters; the scatter plots in the lower left part illustrate the actual bivariate distributions between the parameters. The highlighted $\del{\lambda,\nu,\eta}$ (yellow) and $\del{\zeta,\sigma_Q,\sigma_W}$ (green) parameters stand for quantities that show clear correlation patterns among themselves. The yellow subspace seems to be independent of its green counterpart.}
\label{fig:imperfect_posterior}
\end{figure}

In Figure~\ref{fig:imperfect_posterior} the posterior distribution is visualized in terms of \textit{iid} samples. The clear correlation pattern between the \textit{QoI} indicates that the scale parameter $\zeta$ of the log-normal distribution describing the breach erosion, i.e. quantifying the uncertainty in the global scaling coefficient, is almost independent of other model parameters $\del{\lambda,\nu,\eta}$. They show considerable correlation coefficients among themselves, with $\rho_{\lambda,\nu} = -0.95$ being the most distinct correlation close to linear dependency. On the one hand it indicates an over-parametrization, i.e. one of the two parameters could be neglected in the model formulation without losing information or abandoning physics. On the other hand correlations are commonly caused by data that is not informative enough. The prior correlation between the erosion formula exponents $\rho_{\nu,\eta} = -0.1$, from the epistemic point of view, changed to a more noticeable relation in the posterior with $\rho_{\nu,\eta} = -0.69$. Physically speaking, the dependency between highly intense sediment transport and the impact of cross-sectional shape is more accentuated. Looking at the marginal distributions in the diagonals of Figure~\ref{fig:imperfect_posterior}, the inferential information contained in the data used for the Bayesian update (see Table~\ref{tbl:data}) is evident. The non-informative priors $\pi_{\Lambda}$ and $\pi_{Z}$ changed to peaky shaped distributions. Otherwise the marginal posteriors of exponents $\nu$ and $\eta$ are not far from their prior $\pi_{\bm{M}}\del{\bm{m}}$ in Eq.~\eqref{eq:prior_m}. For $\eta$ the mean value is slightly shifted to a more negative value, whereas $\mu_{\nu}$ did not change significantly, but $\sigma_{\nu}$ decreased notably. The marginal distribution of the scale parameter $\zeta$ shows a wide plateau close to zero. A second correlation pattern is noticed between the parameters $\del{\zeta,\sigma_Q,\sigma_W}$ which describe uncertainties of the dam breach model. The lower the uncertainty in erosion quantification is, i.e. lower values of $\zeta$, the higher will be the values of structural model and data uncertainties $\sigma_Q$ and $\sigma_W$. This additional pattern seems to be nearly independent of the correlations between parameters $\del{\lambda,\nu,\eta}$ that characterize the dam breach erosion but not the associated uncertainties. Estimating the most probable values of the joint posterior distribution for all \textit{QoI}, roughly yields $\del{\hat{\lambda},\hat{\zeta},\hat{\nu},\hat{\eta},\hat{\sigma_Q},\hat{\sigma_W}} = \del{-8.4,0.34,4.1,-0.61,0.22,0.14}$ (see Table~\ref{tbl:mode}).

\subsection{Goodness of Fit}

\begin{figure}[t]
\centerline{\includegraphics[height=3.7in]{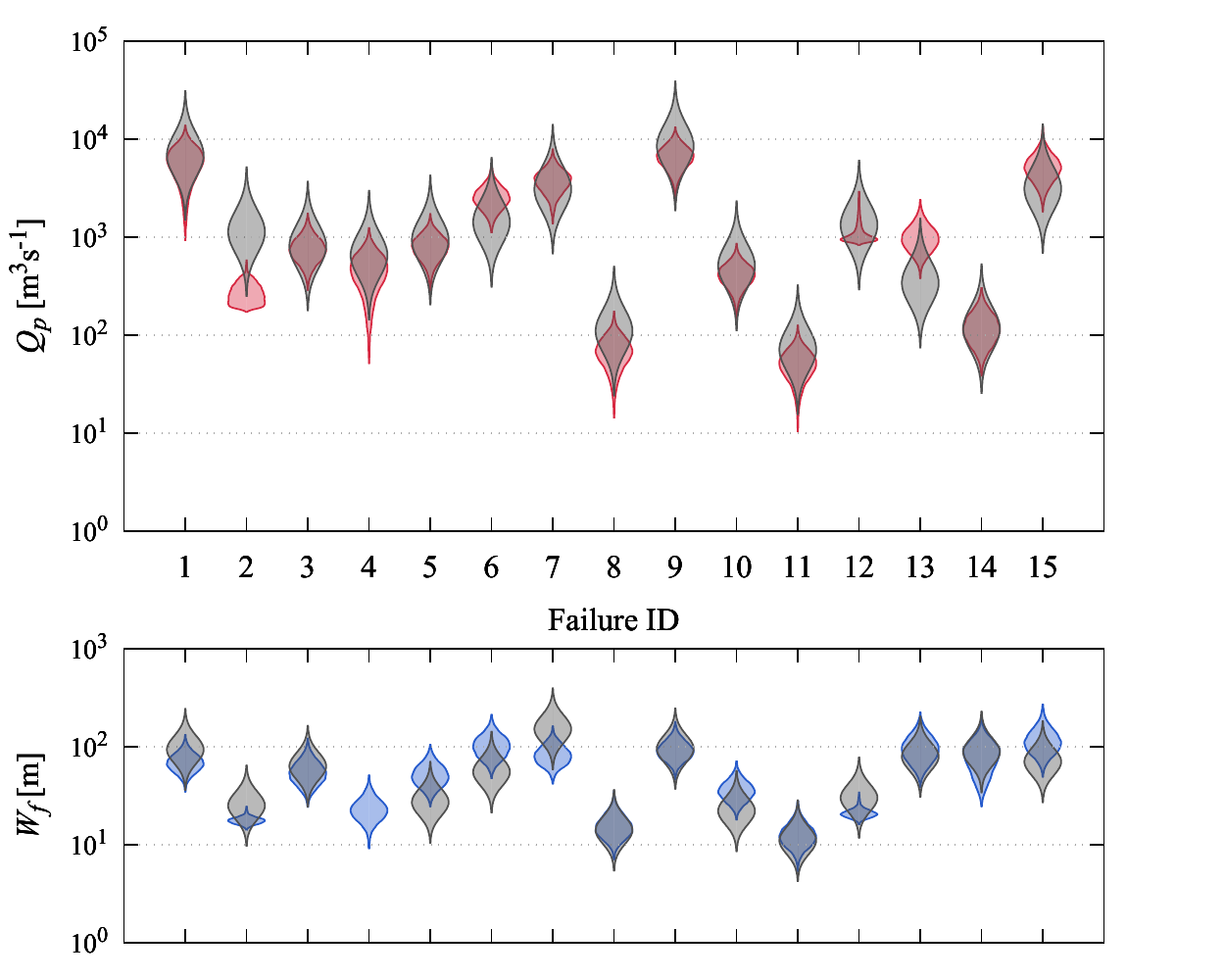}}
\caption{Performance evaluation of model inversion: comparing observed data including calibrated residual uncertainties (gray violins) with modeled data (colored violins). The 15 historic dam failure events used for model inversion (see Table~\ref{tbl:data}) are shown in the upper plot containing peak discharge $Q_p$ data (red) and final breach width $W_f$ data (blue), respectively, in the lower plot.}
\label{fig:imperfect_violins}
\end{figure}

\begin{figure}[t]
\centerline{\includegraphics[height=2in]{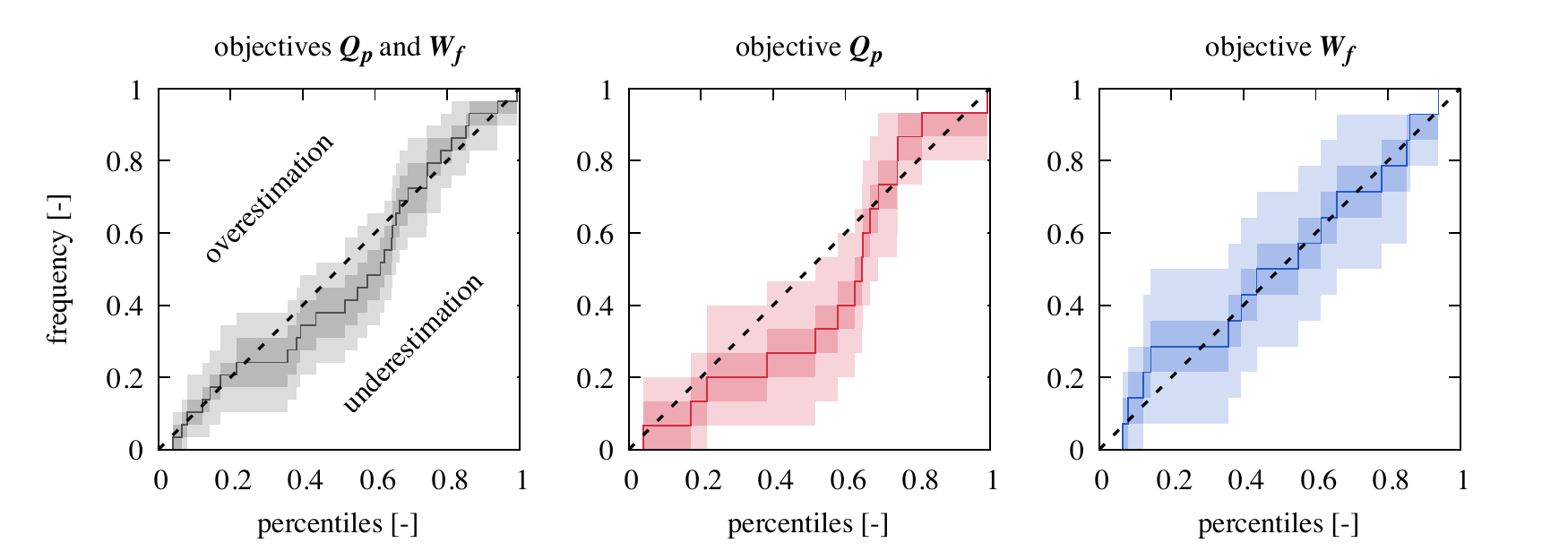}}
\caption{Percentile plot of model inversion performance evaluation, from left to right: summing up both objectives $Q_p$ and $W_f$ (grey); and splitting objectives peak discharge (red) and breach width (blue). Uncertainty bounds were estimated through sampling with replacement, also referred to as bootstrapping \citep{Efron_1979}: median value (line), $\sbr{25,75}\%$ interval (dark area), and $\sbr{5,95}\%$ interval (light area).}
\label{fig:percentiles}
\end{figure}

Running the dam breach model with data used for model inversion and estimated most probable values for unknown parameters, raises the possibility to evaluate the performance of the model inversion. In Figure~\ref{fig:imperfect_violins} both objectives $Q_p$ and $W_f$ are displayed. Observed data of peak discharge and final breach width spread over two and one orders of magnitude, respectively, represented by probabilistic means (violins) because of superimposed residual uncertainties $\sigma_Q$ and $\sigma_W$. In many cases an almost complete overlap of the violins based on modeled and observed data is evident. Otherwise large parts of the violin tails show agreement, i.e. the model inversion appears to have succeeded. A quantitative summary of inversion performance is given in Table~\ref{tbl:error}. The average inversion error amounts to $\widehat{\textrm{E}}\sbr{\bm{r}}=-0.03\pm0.04$ and implies no biased performance as expected. Analyzing the objectives separately shows small but not significant biases $\widehat{\textrm{E}}\sbr{r_Q}=-0.06\pm0.06$ and $\widehat{\textrm{E}}\sbr{r_W}=0.01\pm0.04$, respectively. This insight is disclosed visually in Figure~\ref{fig:percentiles}. Final breach width tends to be overestimated, whereas peak discharge predictions tend towards underestimation in the mid percentile range and overestimation in the upper percentile range. Considering both objectives, as done during the process of model inversion, these trends diminish over the full range of percentiles. The average $95\%$ prediction interval is approximately $0.4$ order of magnitude when considering both objective quantities. Evaluating only peak discharge yields $\widehat{\textrm{I}}_{95}\sbr{r_Q}\approx0.5$, being within the known range for peak discharge predictions. A clearly smaller interval is observed when looking exclusively at breach width $\widehat{\textrm{I}}_{95}\sbr{r_W}\approx0.35$ compared to the literature \citep{Wahl_2004}. Further, the correlation measure between GoF statistics $\widehat{\rho}\sbr{r_Q,r_W}\approx0.2$ is fairly small. This indicates that even the simple residual model in Section~\ref{sec:residual} is sufficiently adequate in the case at hand. The results are justified by the assumption of a residual model where only the residual widths of the model output are investigated but no inference about their correlation is allowed for. Considering the variances of error sources (see Table~\ref{tbl:error}) the inference analysis indicates that about $\nicefrac{2}{3}$ of the total variance arises from residual uncertainty $\widehat{Var}\sbr{\bm{\epsilon}}$ whereas the remaining part seems to originate from parametric uncertainty $\widehat{Var}\sbr{\bm{y}-\tilde{\bm{y}}}$.

\subsection{Alternative of Zero Noise Assumption}\label{sec:perfect}

The results of the successful inference analysis in the previous Section is not directly applicable for hydrograph predictions of a potentially failing dam. The reasons are twofold. On the one hand the observed data $Q_p$ and $W_f$ used for model calibration is not equivalent to the main output of the computational model, which is the outflow hydrograph used as input for flood wave propagation models. This ambiguity restricts potential model errors to be incorporated into the dam breach model for hydrograph predictions. On the other hand the simple residual model applied in this study (see Eq.~\eqref{eq:sigma_epsilon}) prevents to distinguish between structural model errors measurement errors, both represented as part of the residuals $\bm{\epsilon}$. Consequently neglecting the residuals when predicting the hydrograph, for reasons mentioned before, will likely lead to an underestimation of the model uncertainties and the reliability of model predictions is at stake.

The correlation pattern observed in the posterior distribution (see Figure~\ref{fig:imperfect_posterior}) suggests that changes in the residual model presumably will not affect the set of model parameters $\del{\lambda,\nu,\eta}$, but will have an impact on $\zeta$. This represents a quite particular case, while the parameters are usually affected by a change of the residual model \citep[e.g.][]{Thyer_2009}. Due to the modularity concept of the Bayesian multilevel framework \citep{Nagel_2015}, a change of the residual model does not considerably impinge on the other modules from a methodological and technical point of view. Thus, to quantify the parameter uncertainties for direct and predictive model application, a further simplification in the residual model is constituted.

The previous residual model in Eq.~\eqref{eq:sigma_epsilon} is based on the strong assumptions of additivity, homoscedasticity, and Gaussianity. Now these assumptions are substituted by assuming ``perfect'' data and model. Hence, all imperfections
\begin{equation}
\Vert \bm{\Sigma_\epsilon} \Vert \rightarrow \bm{0}
\label{eq:sigma_zero}
\end{equation}
are expected to diminish, i.e. their variability is not bounded to the assumptions of additivity, homoscedasticity, and Gaussianity anymore, but will be compensated by the variability of model parameters. This residual model is referred to as \textit{zero noise limit} hereinafter. Accordingly the \textit{QoI} are reduced to $\bm{\phi}=\del{\lambda,\zeta,\nu,\eta}$ representing a subset of the residual calibration. Now the \textit{QoI} comprise the hyper-parameters $\del{\lambda,\zeta}$ of the log-normal distribution for $\gamma_i$ and the exponents $\nu$ and $\eta$, quantifying the sediment transport within the breach.

The assumption of $\Vert \bm{\Sigma_\epsilon} \Vert \rightarrow \bm{0}$, i.e. noise-free measurements and a perfectly accurate forward model, is obviously far from the true representation of $\bm{\epsilon}$ in case of dam breach data, but no information about the level of measurement errors is at hand and potential structural errors will be compensated by parametric uncertainties in $\zeta$. Variability in the data is still present due to varying inputs across several experiments. To incorporate the change of the residual model in Eq.~\eqref{eq:sigma_zero} into the framework outlined in Section~\ref{sec:calibration}, the estimation of the integrated likelihood formulation (see Eq.~\eqref{eq:likelihood_int}) is slightly modified and defined as the kernel density function
\begin{equation}
\hat{f}\del{\bm{y}_i\mid\bm{m},\bm{\theta_u}} = \frac{1}{K}\sum\limits_{k=1}^{K}\mathcal{K}\del{\bm{\epsilon}_i^{\del{k}}},
\end{equation}
where $\mathcal{K}\del{\bm{\epsilon}_i^{\del{k}}}$ are Gaussian kernels and the bandwidth selection is based on the normal distribution approximation \citep{Silverman_1986}.

\begin{figure}[t]
\centerline{\includegraphics[height=3in]{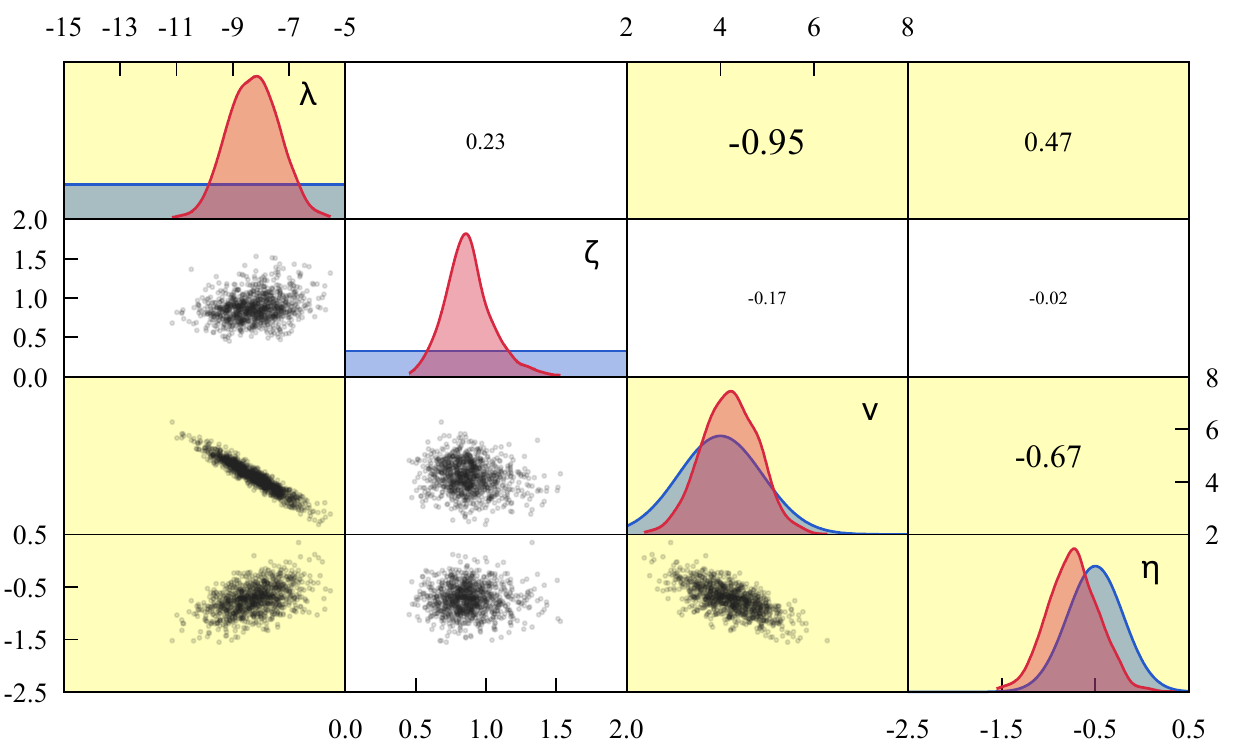}}
\caption{Posterior distribution samples, assuming zero noise limit: illustration is similar to Figure~\ref{fig:imperfect_posterior}. The correlation pattern of parameters $\del{\lambda,\nu,\eta}$ (yellow) remains, whereas the model parameter $\zeta$ is independent.}
\label{fig:perfect_posterior}
\end{figure}

This calibration configuration of the zero noise limit is seen as an alternative to the residual calibration. Reusing the data in Table~\ref{tbl:data} is legitimate as long as the knowledge inferred from residual calibration is not applied. Therefore the prior distribution defined in Section~\ref{sec:priors} is not changed, except from neglecting $\pi_{\bm{E}}\del{\bm{\Sigma_\epsilon}}$. The performance of the MCMC simulation can be found in Appendix~\ref{app:mcmc_perf_A}. The posterior resulting distribution is illustrated in Figure~\ref{fig:perfect_posterior}. Neither the correlation pattern nor the shape of the marginals of parameters $\del{\lambda,\nu,\eta}$ have changed compared to the residual calibration. The most probable parameter values $\del{\hat{\lambda},\hat{\nu},\hat{\eta}} = \del{-8.3,4.2,-0.67}$ show negligible differences (see Table~\ref{tbl:mode}). A major difference is observed in the marginal distribution of the scale parameter $\zeta$. Now a distinct peak is noticed around $\hat{\zeta}=0.83$, suggesting that the uncertainty in the scaling coefficient increased, compensating for missing residual variability, not only for structural model errors, but for measurement errors as well.

Table~\ref{tbl:error} shows the average expected residual is approximately zero. The average $95\%$ prediction interval is approximately half order of magnitude, slightly more than in case residual calibration. $\widehat{\textrm{I}}_{95}\sbr{r_W}\approx0.4$ is significantly larger, suffering most from the increased correlation measure $\widehat{\rho}\sbr{r_Q,r_W}\approx0.8$. The GoF statistics $r_Q$ and $r_W$ are strongly correlated and therefore the residuals $\bm{r}$ do not show properties of white noise. These correlated errors are attributed to either global model inadequacies and/or data inconsistencies, caused by the simplification of the zero noise assumption.

\subsection{Comparison of Residual Models}

Comparing the posterior distributions in Figure~\ref{fig:imperfect_posterior} and Figure~\ref{fig:perfect_posterior} clearly illustrates, that on the one hand the characterization of the sediment transport formula within the breach (see Eq.~\eqref{eq:qs}) is independent of the residual model approach. The correlation pattern among the parameters $\del{\lambda,\nu,\eta}$ as well as the set of their most probable values is not dependent on the choice of the residual model. This can be interpreted as a plausibility check for the resulting posteriors. Taking the median value $e^{\hat{\lambda}}$ of the log-normal distribution describing the stochastic scaling coefficient $\gamma$ and writing the transport formula in a deterministic way, one finds $q_s \approx 0.00025 \cdot v^{4.15} \cdot r_{hy}^{-0.65}$. The exponents $\nu = 4.15$ and $\eta = -0.65$ are not much different from prior knowledge (see Eq.~\eqref{eq:prior_m}). On the other hand, parameters that contain information about the model uncertainties $\del{\zeta,\sigma_Q,\sigma_W}$ are strongly dependent on the residual model. When assuming uncorrelated residuals, the model and data uncertainty is represented by $\sigma_Q=0.22$ and $\sigma_W=0.14$ and parametric uncertainty describing the scaling coefficient $\gamma$ is quantified by the \textit{QoI} $\zeta=0.34$. For zero noise limit assumption all residual uncertainty is covered by parametric uncertainty of the breach erosion process, consequently $\zeta$ is increased from $0.34$ to $0.83$.

\begin{table}[t]
\caption{Most probable parameter value estimations of the joint posterior distribution including the estimation errors measured through bootstrapping.}
\centering
\begin{tabular}{
	c
	S[table-format=-2.3,table-figures-uncertainty=1]
	S[table-format=-2.3,table-figures-uncertainty=1]}
\toprule
\multirow{2}{*}{\textit{QoI}} & \multicolumn{2}{c}{residual model} \\
\cmidrule{2-3}
& \multicolumn{1}{c}{residual calibration} & \multicolumn{1}{c}{zero noise limit} \\
\midrule 
\(\hat{\lambda}\) & -8.37 \pm 0.06 & -8.25 \pm 0.02 \\
\(\hat{\zeta}\) & 0.340 \pm 0.010 & 0.833 \pm 0.003 \\
\(\hat{\nu}\) & 4.12 \pm 0.04 & 4.17 \pm 0.01 \\
\(\hat{\eta}\) & -0.610 \pm 0.003 & -0.669 \pm 0.006 \\
\(\hat{\sigma}_Q\) & 0.220 \pm 0.003 & \\
\(\hat{\sigma}_W\) & 0.139 \pm 0.002 & \\
\bottomrule
\end{tabular}
\label{tbl:mode}
\end{table}

The differences between the two residual models is evident by performing a variance decomposition of model residuals (see Table~\ref{tbl:error}). When assuming diminishing residuals during inversion, the overall output variance is increased from $0.052$ to $0.057$. No increase is observed in the variance of peak discharge residuals (originally $70\%$ assigned to the data uncertainty parameter $\sigma_Q$). A more prominent increase in variance is detected for final breach width prediction errors, namely, from $0.032$ to $0.042$ (primarily $60\%$ covered by $\sigma_W$). This is explained by the strong correlation of residuals $r_Q$ and $r_W$ for the zero noise assumptions, where model outputs $\tilde{\bm{y}}=\del{\tilde{Q}_p,\tilde{W}_f}$ are correlated due to the physical breach formation processes implemented in the dam breach model. Breaking up this tie during residual calibration shows, that the global uncertainty in $Q_p$ is responsible of overestimating the error in $W_f$ predictions. Furthermore, the residual calibration suggests that more than half of the global uncertainties are associated with residual variability. To assign these global uncertainties to model inadequacies and measurement errors, more sophisticated error modeling would be needed that is beyond the scope of this study, but being a topic for future work. While pure measurement errors are typically closer to white noise, the model inadequacies can be modeled by a functional error term, often represented by Gaussian processes and dependent on the classes of known model parameters $\bm{d}$ and $\bm{z}$ \citep[e.g.][]{Kennedy_2001,Brynjarsdottir_2014}.

\begin{table}[t]
\caption{Error statistics of the model inversion performance evaluation: Estimated mean error, $95\%$ confidence interval, correlation coefficient between objective quantities $Q_p$ and $W_f$, and variances of different error sources.}
\centering
\footnotesize{
\begin{tabular}{
  l
  c
  S[table-format=-1.2,table-figures-uncertainty=1]
  S[table-format=-1.2,table-figures-uncertainty=1]
  S[table-format=-1.2,table-figures-uncertainty=1]
  S[table-format=-1.1,table-figures-uncertainty=1]
  S[table-format=-1.1,table-figures-uncertainty=1]
  c
}
\toprule
\begin{tabular}{@{}l@{}}residual \\ model\end{tabular} &
\begin{tabular}{@{}c@{}}output \\ quantity\end{tabular} &
\(\widehat{\textrm{E}}\sbr{\bm{r}}\) &
\(\widehat{\textrm{I}}_{95}\sbr{\bm{r}}\) &
\(\widehat{\rho}\sbr{r_Q,r_W}\) &
\multicolumn{3}{c}{\begin{tabular}{@{}ccc@{}}
	\(\widehat{Var}\sbr{\bm{r}}\) &
	\(\widehat{Var}\sbr{\bm{y}-\tilde{\bm{y}}}\) &
	\(\widehat{Var}\sbr{\bm{\epsilon}}\) \\
	\cmidrule{1-3}
	\multicolumn{3}{c}{(expressed in \(10^{-2}\))}
\end{tabular}} \\
\midrule
\multirow{3}{*}{\begin{tabular}{@{}l@{}}residual \\ calibration\end{tabular}}
& \(Q_p\)      & -0.06 \pm 0.06 & 0.50 \pm 0.01 &
               &  7.1 \pm 0.3 & 2.2 \pm 0.3 & 4.0 \\
& \(W_f\)      &  0.01 \pm 0.04 & 0.35 \pm 0.01 &
               &  3.2 \pm 0.2 & 1.2 \pm 0.2 & 1.7 \\
& \(Q_p, W_f\) & -0.03 \pm 0.04 & 0.44 \pm 0.02 & 0.20 \pm 0.03
               &  5.2 \pm 0.3 & 1.7 \pm 0.2 & 2.9 \\
\midrule
\multirow{3}{*}{\begin{tabular}{@{}l@{}}zero \\ noise \\ limit\end{tabular}}
& \(Q_p\)      & -0.03 \pm 0.06 & 0.52 \pm 0.03 &
               & 7.1 \pm 0.8 & 7.1 \pm 0.8 & 0.0 \\
& \(W_f\)      & 0.05 \pm 0.04 & 0.41 \pm 0.02 &
               & 4.2 \pm 0.3 & 4.2 \pm 0.3 & 0.0 \\
& \(Q_p, W_f\) & 0.01 \pm 0.04 & 0.47 \pm 0.02 & 0.84 \pm 0.04
               & 5.7 \pm 0.5 & 5.7 \pm 0.5 & 0.0 \\
\bottomrule
\end{tabular}
}
\label{tbl:error}
\end{table}

To qualitatively compare the two calibrated dam breach models, differing in the underlying residual model, and selecting the right one is known under Bayesian model selection \citep{Schoeniger_2014}. This has not been accomplished in this study due to the computational cost of evaluating the evidence $C$ (see Eq.~\eqref{eq:posterior}). The motivation to perform the non-standard alternative of zero noise limit (see Section~\ref{sec:perfect}) is crucial when facing a real case application where reliable hydrograph predictions are requested.

\section{Discussion of Model Application}\label{sec:discussion}

The probabilistic modeling framework and the according calibration results have been discussed in the previous section. Here the applicability of the resulting probabilistic dam breach model is discussed by applying the proposed model in a test case.

The problem statement of a hypothetical dam failure originates from a numerical benchmark organized by the International Commission on Large Dams \textit{ICOLD} that took place in Graz 2013 \citep{Graz_2013}. The parameter definition is given as follows (compare with Table~\ref{tbl:parameters}). The embankment dam is characterized by dam height $h_d = \SI{61}{\meter}$, crest width $w_c = \SI{24}{\meter}$, and embankment slope $s_e = 3$. The reservoir is described by a drop in reservoir level $\Delta H_r = h_d = \SI{61}{\meter}$ and according release of water volume $\Delta V_r = \SI{38276344}{\cubic\meter}$, and the reservoir basin shape parameter approximately follows a uniform distribution $\alpha \sim U\del{2.5,3.2}$ (estimated from retention curve). The breach side angle at the top is assumed to be uniformly distributed $\beta \sim U\del{50,85}$, the maximal breach height is $\Delta H_b = \SI{61}{\meter}$, and the initial breach depth is assumed to be $Y_{b,0} = \SI{50}{\meter}$, and therefore $r_0 = 0.82$. The erosion of dam material is quantified by the transport formula in Eq.~\eqref{eq:qs}. For reasons of applicability and reliability discussed in Section~\ref{sec:perfect} the calibration results under the assumption of zero noise are considered here. Making use of the most probable values as point estimates (see Table~\ref{tbl:mode}), the applied transport formula yields $q_s \approx \gamma \cdot v^{4.2} \cdot r_{hy}^{-0.67}\,\si{\square\meter\per\second}$ where $\gamma \sim LN\del{-8.3,0.83}$.

\begin{figure}[t]
\centerline{\includegraphics[height=3.2in]{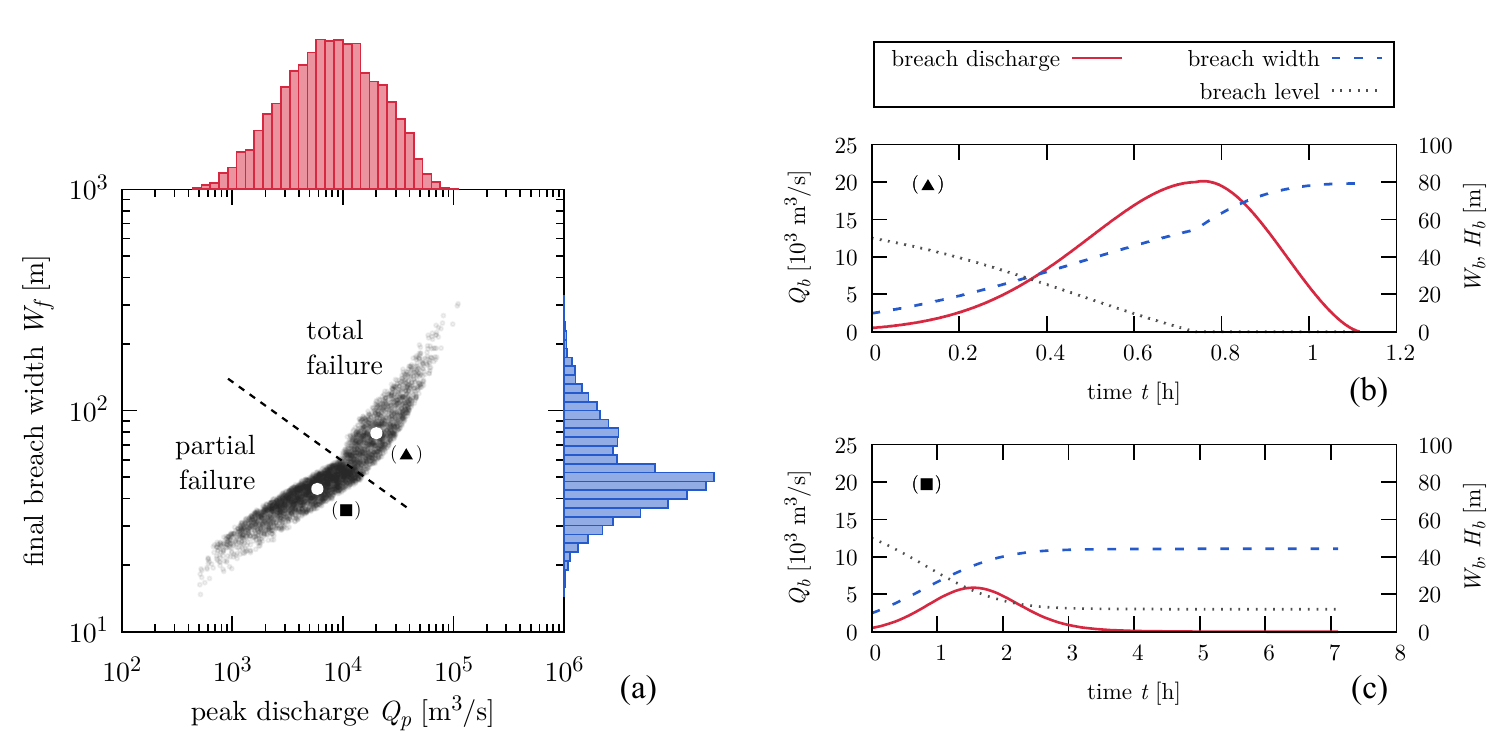}}
\caption{Application example of proposed probabilistic dam breach model (input data taken from \citet{Graz_2013}). Dam breach hydrograph predictions of hypothetical embankment dam by modeling physical breach formation processes: 5000 model evaluations and resulting peak discharge $Q_p$ and final breach width $W_f$ and their marginal distributions (a); and explicitly showing time series of two different samples, indicated with ($\blacksquare$) and ($\blacktriangle$), for breach outflow, breach width, and breach bottom level ((b) and (c)).}
\label{fig:ICOLD_app}
\end{figure}

Running a Monte-Carlo simulation of 5000 model evaluations leads to 5000 different progressive breach formation processes. Here a Latin Hypercube Sampling (LHS) was chosen as sampling strategy for the purpose of faster convergence of the MC simulation. In Figure~\ref{fig:ICOLD_app} the resulting distribution of the hydrograph's peak discharge and the final breach width are shown. Two distinct behaviors can be observed in Figure~\ref{fig:ICOLD_app}a: parameter sets with high erosion rates (large $\gamma$ parameter realization) lead to total failure of the dam, indicated by final breach $W_f$ being larger than dam height $h_d = \SI{61}{\meter}$, whereas parameter sets with low erosion rates (small $\gamma$ parameter realization) lead to a partial failure only. In this exemplary application the driving force of the reservoir water does not seem to be sufficient to erode the large body of dam material in many parameter combinations. Two representative hydrographs for complete and partial failure are shown in Figure~\ref{fig:ICOLD_app}. In case of total failure the progression is obvious and peak breach outflow is reached within less than $\SI{1}{\hour}$, nearly at the same time when the breach bottom hits the fixed bottom, i.e. transition from vertical erosion to lateral widening. In case of partial failure the gradual erosion is not fast enough to initiate a progressive failure mechanism and the breach bottom never reaches the fixed bottom. Here the duration of the failure process is much longer (a few hours).

The performance of the breach model proposed in this study is based on simple physical assumptions. Thus the model behavior is assumed to be close to physical processes of real dam failures, including the related probabilities of observing one or the other formation process. The resulting peak discharge distribution (see Figure~\ref{fig:ICOLD_app}a) is therefore much wider than the predicted peak discharges of the benchmark participants (ranging from $10$ to $\SI{40e3}{\cubic\meter\per\second}$). Neglecting the possibility of partial failure in this case might lead to a significant overestimation of the predicted hydrograph. In this case, the consequent flood wave calculation would result in too intensive hydraulic impact, hence the according quantification of the dam break risk is biased.


\section{Conclusions}\label{sec:conclusions}

Real dam break risk analysis requires for reliable breach models capable to predict possible breach outflow hydrographs. The challenge to predict a worst case scenario, that still is regarded as a physically feasible event, is of particular interest for engineering applications. By contrast, seeking for more accurate breach models often loses sight of reliably predicting less probable events. In this study the lack of knowledge in describing the progressive process accurately is compensated by the development of a probabilistic dam breach model framework where uncertainties on different levels are quantified. The focus is on homogeneous and non-cohesive earthen embankment dams whose potential failures need to be analyzed for the purpose of further flood risk assessment.

Uncertainties in the quantification of dam material erosion is used to tune the newly proposed dam breach model in a Bayesian fashion. Prior information about dam breach model parameters together with data from historical dam failure events are used to perform Bayesian inference. Model inversion is accomplished by sampling from the posterior distribution through DE-MCMC simulations. The convergence to the posterior distribution and the consequent statistical inference of the multilevel model was successful using the information contained in a data set of 15 real dam failure events. The very same data has been used to carry out deterministic model calibration of simplified physics-based dam breach models comparable to the model developed in the study at hand \citep[e.g.][]{Macchione_2008a}. By re-using the same data the focus is on the proposed framework itself instead of possible changes in prediction quality due to different data.

Observed data of peak discharge $Q_p$ and final breach width $W_f$ are used as output quantities. Output imperfections, consisting of structural model errors and measurement errors, are defined as residuals and their prior knowledge is represented by independent Gaussians for each output quantity. Inferring for parametric and residual uncertainties results in a posterior distribution with a distinct correlation pattern. Model parameters that characterize the sediment transport rate inside the breach are independent of residual uncertainty. Moreover, they are not related to the hyper-parameter describing the aleatory uncertainty of the global scaling coefficient $\gamma$. The characterization of the average sediment transport can be best described by $q_s \approx 0.00025 \cdot v^{4.15} \cdot r_{hy}^{-0.65}\,\si{\square\meter\per\second}$. Comparing this formulation with the theoretical values from the transport law by Meyer-Peter \& M{\"u}ller and the friction law by Manning ($\nu = 3.0$ and $\eta = -0.5$), the transport within the breach can be attributed to total-load-like ($\nu > 3.0$) and more sensitive to cross-sectional geometries ($\eta < -0.5$). However, the variability of $q_s$, described by the hyper-parameter $\zeta$, is dependent on the variability of the residuals. The latter are correlated among themselves in addition.

Assessing the goodness-of-fit (GoF) statistics of the performed model inversion yields a $95\%$ confidence interval for model predictions around half orders of magnitude for the model output peak discharge $Q_p$, falling in the range of values reported in literature for other dam breach prediction tools \citep{Wahl_2004,Froehlich_2016}. The model output final breach width $W_f$ shows a $95\%$confidence interval around third orders of magnitude. The variability of the GoF statistics is explained by roughly $60\%$ from the residual variability and $40\%$ covered by the different parametric uncertainties.

The quantities $Q_p$ and $W_f$ representing the observed data are not conforming with the final model output in real model applications. It is the main motivation of the proposed probabilistic dam breach model to predict the time series of breach outflow (breach hydrograph) needed as upper boundary condition in state-of-the-art flood routing models. Assuming non-zero residuals and consequently incorporating reversely the estimated residuals of $Q_p$ into predicted hydrographs is impractical. The clear correlation pattern in the posterior distribution and the hierarchical character of the Bayesian multilevel framework allows for an alternative formulation of the residual model, where residual uncertainties are assumed to diminish. This \textit{zero noise limit} represents a subspace of the full problem. The inference results in this unusual setup show, that the characterization of $q_s$ does not change, but uncertainty due to both structural model errors and measurement errors is compensated by parametric uncertainty of the sediment transport rate quantification. As a result the uncertainty of model predictions is overestimated, whereas it is probably underestimated in the case of full residual calibration. On these grounds the zero noise assumption is still favorable when applying the proposed modeling framework to a real case, when the computational model predicts full hydrographs instead of maximum flows only. To avoid the non-standard residual model \textit{zero noise limit}, two additional ingredients are needed in the proposed modeling framework: (i) improved error modeling, allowing for disentangling structural model errors and measurement errors; and (ii) development of a physically based technique to reversely apply uncertainties of $Q_p$ and $W_f$ to predicted hydrographs. In fact the community is encouraged to quantify and/or reduce the uncertainties contained in the observed data and thereby improving the precision of the model substantially \citep{Morris_2008,ASCE_2011}.

The probabilistic dam breach model was successfully applied to a test case, which consists of a hypothetical embankment dam \citep{Graz_2013}. It clearly demonstrated the benefits of the proposed modeling framework, when taking into account the parametric uncertainties that were quantified in this study. The simplified model, which allows for physical interpretation of the results, combined with the motivation of improving prediction reliability, enabled by enhanced methodologies, provides an adequate compromise between model accuracy and model reliability.

Improving the formulation of the deterministic and simplified dam breach model, such as introducing additional physical parameters (e.g. discharge coefficient), requires to re-assess all parametric uncertainties by inference analysis. To strengthen and/or critically assess the proposed dam breach model framework, the same inferential analysis can be performed subsequently by inferring from an additional data set of observed data, where the prior is defined by the posterior of this study, as done in \cite{Peter_2017}. Since the validity of similar dam breach model formulations have been proven \citep[e.g.][]{Chinnarasri_2009,Capart_2013,Vonwiller_2015}, observed data from real dam failure events is regarded to provide its maximum value within the proposed calibration procedure to further improve the empirical knowledge of model configurations. The gathering of dam failure data containing qualitatively and quantitatively enough information remains a main challenge.

Eventually, to deal with a probabilistic breach outflow hydrograph, the need for fast and accurate flood wave calculation tools is of paramount importance. Recent advances in computer sciences open a new door to fill this gap \citep{Kalyanapu_2011,Lacasta_2014,Reguly_2015}. The possibility of running two dimensional flood wave calculation as Monte-Carlo simulation with random hydrograph is real \citep{Peter_2017}. The information contained in the resulting probabilistic flood maps is abundant. Finally, there is a strong need for engineering design aids and standard procedures for embankment breach analysis and hazard management that quantitatively incorporate uncertainties.

\appendix

\section{Dam breach model initial conditions}\label{app:IC}

Fixed breach bottom $H_{b,min}$, initial breach level $H_{b,0}$, initial reservoir level $H_{r,0}$ and corresponding reservoir volume $V_{r,0}$ are calculated through (see Figure~\ref{fig:sketches})

\begin{subequations}
\begin{align}
H_{b,min} &= h_{d} - \Delta H_b \\
H_{b,0} &= H_{b,min} + \del{1 - r_0} \Delta H_r \\
H_{r,0} &= H_{b,min} + \Delta H_r \\
V_{r,0} &= \Delta V_r \frac{{H_{r,0}}^\alpha}{{H_{r,0}}^\alpha-{H_{b,min}}^\alpha}.
\end{align}
\label{eq:init}
\end{subequations}

Furthermore, having the breach side angle $\beta$, that is a proxy for the embankment material property, and the initial breach level $H_{b,0}$ is not sufficient. An initial value of the breach top width $W_b$ is needed in addition. In case of triangular breach shape ($\beta=45^{\circ}$) the initial breach width $W_{b,0} = 2h_b$ is unique and the according breach discharge
\begin{equation}
Q_{b,ref} = \sqrt{\frac{512}{3125}g{H_e}^5}
\label{eq:qref}
\end{equation}
is taken as reference value. However, $W_{b,0}$ is not unique for all other choices of $\beta$. Thus, the restriction is set, that the initial breach discharge $Q_b\del{\beta,W_{b,0}}$ is equal to initial discharge of a triangular breach in Eq.~\eqref{eq:qref}) \citep{Franca_2005}. When assuming that the reservoir is filled to capacity ($H_{r,0} \mathrel{\widehat{=}} h_d$), the initial breach width can then be approximated as
\begin{equation}
W_{b,0}\del{\beta} \approx
\frac{16}{25}\del{5-\frac{\beta}{24}}\del{h_d-H_{b,0}}.
\label{eq:init_width_approx}
\end{equation}

\section{Empirical transport and friction formulas}\label{app:transport}

The exponents $\nu$ and $\eta$ in Eq.~\eqref{eq:qs} vary strongly depending on the field of application. Existing transport, respectively friction quantification tools, are based on laboratory and/or field test data that is further processed to result in simple empirical formulas. Numerous examples of such formulas can be found in the literature. According to \citet{Wu_2008} and \citet{Machiels_2011} lower and upper bounds for the exponents $c_{1 \ldots 4}$ (see Section~\ref{sec:erosion}) are assumed in this study as
\begin{subequations}
\begin{align}
c_1 & \in \intcc{1.0,2.2} \\
c_2 & \in \intcc{0.0,2.0} \\
c_3 & \in \intcc{-0.6,0.0} \\
c_4 & \in \intcc{0.4,0.72}.
\end{align}
\end{subequations}
Uniformly sampling from the above ranges leads to an empirical approximation of possible model parameter combinations $\nu = 2c_1+c_2$ and $\eta = c_1\del{1-2c_4}+c_3$.

\section{Markov Chain Monte Carlo (MCMC) algorithm}\label{app:mcmc_alg}

The MCMC algorithm applied in this study \citep{Braak_2006} runs $N$ Markov chains in parallel and has been proven to ensure both detailed balance and ergodicity. Modifications with improved efficiency for high dimensional posterior are available in \citet{Vrugt_2009a} or \citet{Laloy_2012}. The heart of this MCMC algorithm is the absence of a predefined and fixed proposal distribution, but new parameter configurations are instead proposed in an adaptive manner for $j$'th chain as
\begin{equation}
\bm{\phi}_p = \bm{\phi}_j + \delta \del{\bm{\phi}_{r2} - \bm{\phi}_{r1}} + \bm{e},
\end{equation}
with $\bm{\phi}_j$ being the actual parameters of chain $j$, $r1$ and $r2$ are randomly selected chains different from $j$, $\delta = 2.38 / \sqrt{2d}$ is a scaling factor for the jumping width to ensure efficient acceptance rates. To ensure ergodicity of the Markov Chain, $\bm{e}$ is drawn from a symmetric distribution with a small variance compared to that of the target, but with unbounded support. Initial states are generated by sampling from the prior distribution $\pi\del{\bm{\phi}}$. In addition, classic Metropolis algorithm is applied, i.e. accepting proposed parameters $\bm{\phi}_p$ with probability
\begin{equation}
a = \textrm{min}\del{1,\frac{\pi\del{\bm{\phi}_p \mid \langle \bm{y}_i \rangle}}{\pi\del{\bm{\phi}_j \mid \langle \bm{y}_i \rangle}}}.
\end{equation}

\section{MCMC performance}\label{app:mcmc_perf}

MCMC simulations run in this study showed time series with acceptance rates between $0.25$ and $0.30$ over the total number of iterations $I$, ensuring good mixing of the random walk and therefore converging optimally to the stationary posterior distribution \citep{Roberts_2001}. Starting from initial states, samples of $I_b$ iterations are rejected until the stationary target distribution was reached, also known as burn-in period. To monitor convergence of MCMC to the target distribution, the diagnostic \emph{potential scale reduction factor PSRF} was applied \citep{Brooks_1998}, in which the variances (covariances in case of multivariate analysis) within each MCMC-chain and between the $N$-chains are compared. Convergence is assumed to be reached if $\textit{PSRF}<1.1$ and the effective number of iterations is $I_{\textit{eff}}=I-I_b$. Samples of $I_{\textit{eff}}$ show strong autocorrelation because of the Markov Chain generation mechanism. The number of effective samples, that is samples showing properties of \textit{iid} (independent and identically distributed) random samples, is estimated through
\begin{equation}
\hat{N}_{\textit{eff}} = \frac{N \cdot I_{\textit{eff}}}{1+2\sum_{l=1}^{L}\hat{\rho}_l},
\label{eq:neff}
\end{equation}
where $\hat{\rho}_l$ is the sample autocorrelation with lag $l$ and maximum lag $L$ is the first odd integer for which $\hat{\rho}_{L+1} + \hat{\rho}_{L+2} < 0$ \citep{Gelman_2014}. MCMC simulation runs until $\hat{N}_{\textit{eff}}>1000$ for each \textit{QoI}. The denominator in Eq.~\eqref{eq:neff} is applied as thinning lag $t_l$ to get rid of the autocorrelation. Finally, the remaining samples represent independent samples of the joint posterior distribution that will undergo further analysis.

\subsection{Residual Calibration}\label{app:mcmc_perf_A}

\begin{figure}[t]
\centerline{\includegraphics[height=4in]{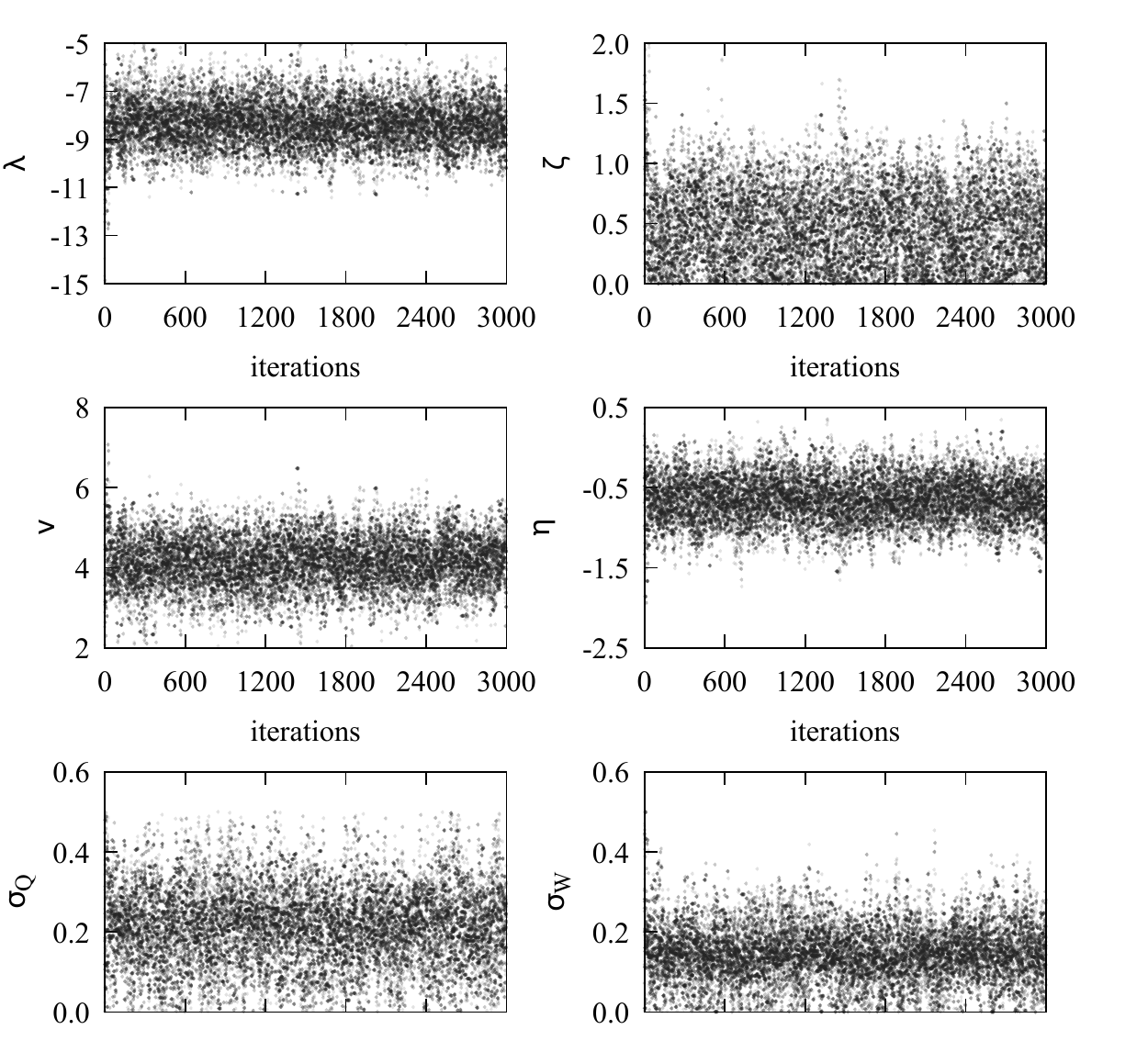}}
\caption{Time Series of the Markov Chain Monte Carlo simulation (differential evolution algorithm with $12$ parallel chains \citep{Braak_2006}) for the quantities of interest $\bm{\phi}=\del{\lambda,\zeta,\nu,\eta,\sigma_Q,\sigma_W}$.}
\label{fig:timeseries}
\end{figure}

\begin{figure}[t]
\centerline{\includegraphics[height=2.5in]{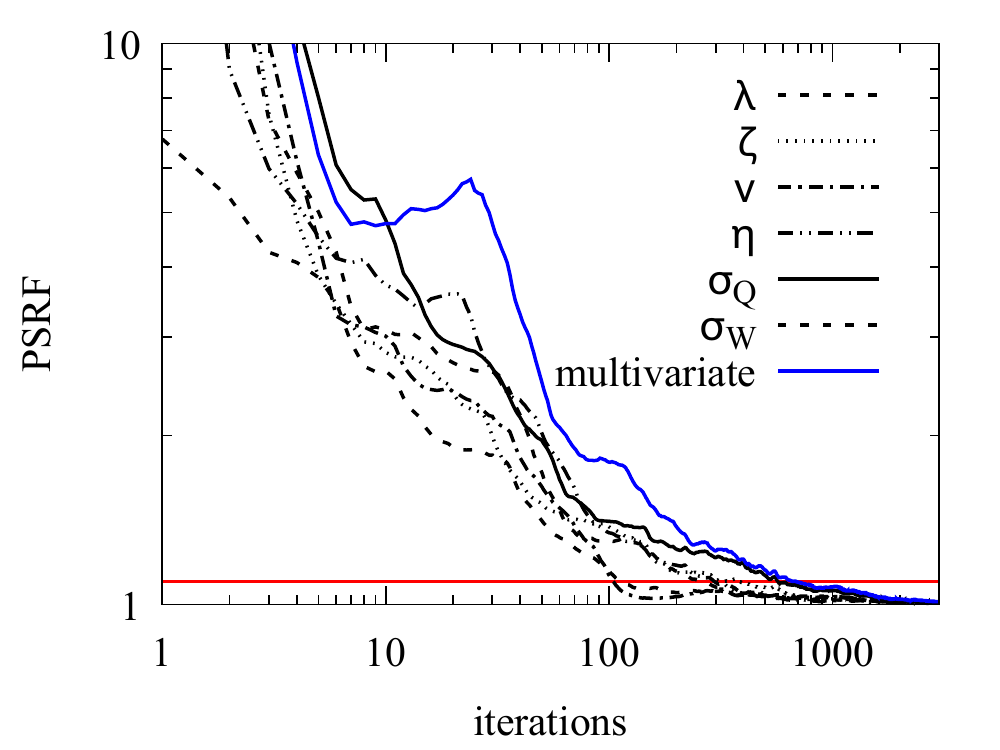}}
\caption{Convergence diagnostic after \citet{Brooks_1998} for the four \textit{QoI} and their multivariate version: after $I_b=800$ iterations all parallel chains in MCMC simulation seem to have reached the stationary target distribution (potential scale reduction factor $\textit{PSRF}<1.1$).}
\label{fig:convergence}
\end{figure}

Running the MCMC simulation with $N=2d=12$ parallel chains results in the time series given in Figure~\ref{fig:timeseries} obtained using a total of $I=3000$ iterations. The time series plot does not show any pattern of bad mixing what is congruent with a mean acceptance rate over all chains of $0.29$. The convergence to a stationary distribution is illustrated by the potential scale reduction factor $\textit{PSRF}$ in Figure~\ref{fig:convergence}. After $300$ iterations the chains are already close to their stationary distribution. The last quantity to reach the limit of $\textit{PSRF}<1.1$ is the multivariate diagnostic and $\sigma_Q$ at $I_b=800$ iterations. After having rejected iterations of the burn-in period, the remaining $2200$ iterations are thinned with $t_l=26$ (see Eq.~\eqref{eq:neff}) to finally obtain approximately $1000$ \textit{iid} posterior samples put together from the $N$ parallel chains.

\subsection{Zero Noise Limit}\label{app:mcmc_perf_B}

Compared to the residual calibration the number of chains $N=12$ was not changed. The mean acceptance rate of $0.28$ is still within the known range of optimal convergence behavior. Due to smaller parameter space $d=4$ convergence is slightly faster and the burn-in period $I_b=600$ is shorter accordingly. Likewise the autocorrelation has decreased and a thinning lag $l_t=19$ was applied to gain \textit{iid} samples. The MCMC simulation was run with $I=2200$ iterations to end up with approximately $1000$ posterior samples.


\section*{Acknowledgments}
The authors would like to thank the Federal Office of Energy of Switzerland (SFOE) for their financial support of the project ``Dam Break Analysis under Uncertainty'' and namely M. G\"{u}ell i Pons and G. Darbre of the Section of Supervision and Safety of Dams for their fruitful discussions. The authors further acknowledge the work of Alessandra Eicher, who provided valuable information in the scope of her Master thesis. The data used are listed in the references.


\bibliographystyle{chicago}
%
%
%
%
%

\end{document}